\RequirePackage[T1]{fontenc}
\documentclass[12pt]{article}

\usepackage[height=8.85in,width=6.45in]{geometry}

\usepackage[utf8]{inputenc}
\usepackage{amsmath}
\usepackage{amssymb}
\usepackage{mathtools}
\numberwithin{equation}{section}
\usepackage{slashed}
\usepackage{braket}
\usepackage[svgnames]{xcolor}
\usepackage[colorlinks,citecolor=DarkGreen,linkcolor=FireBrick]{hyperref}
\usepackage{cite}
\usepackage{graphicx}
\usepackage{tikz}
\usepackage{tikz-cd}
\usepackage{times}
\usepackage{courier}
\usepackage{bm}
\usepackage{subfig}

\usepackage{enumitem}

\usepackage[normalem]{ulem}

\usepackage{soul}

\usetikzlibrary{decorations.pathreplacing,decorations.markings}

\tikzset{
	partial ellipse/.style args={#1:#2:#3}{
		insert path={+ (#1:#3) arc (#1:#2:#3)}
	}
}

\tikzset{
	on each segment/.style={
		decorate,
		decoration={
			show path construction,
			moveto code={},
			lineto code={
				\path [#1]
				(\tikzinputsegmentfirst) -- (\tikzinputsegmentlast);
			},
			curveto code={
				\path [#1] (\tikzinputsegmentfirst)
				.. controls
				(\tikzinputsegmentsupporta) and (\tikzinputsegmentsupportb)
				..
				(\tikzinputsegmentlast);
			},
			closepath code={
				\path [#1]
				(\tikzinputsegmentfirst) -- (\tikzinputsegmentlast);
			},
		},
	},
	mid arrow/.style={postaction={decorate,decoration={
				markings,
				mark=at position .5 with {\arrow[#1]{stealth}}
	}}},
}

\usetikzlibrary{shapes.multipart}
\usetikzlibrary{decorations.pathmorphing}
\tikzset{snake it/.style={decorate, decoration=snake}}

\usepackage{xcolor}
\usepackage{mdframed}

\renewenvironment{figure}[1][]{
	\begin{originalfigure}[#1]
		\begin{mdframed}[linecolor=black!0,backgroundcolor=black!1]
		}{
		\end{mdframed}
	\end{originalfigure}
}

\DeclareMathOperator\Hom{Hom}

\DeclareMathOperator\Rep{Rep}

\definecolor{dgreen}{rgb}{0, 0.55, 0}

\def\CA{{\mathcal A}}
\def\CB{{\mathcal B}}
\def\CC{{\mathcal C}}
\def\CD{{\mathcal D}}

\def\CI{{\mathcal I}}

\def\CL{{\mathcal L}}

\def\CN{{\mathcal N}}

\def\CP{{\mathcal P}}

\def\CS{{\mathcal S}}

\def\CX{{\mathcal X}}

\def\CY{{\mathcal Y}}
\def\CZ{{\mathcal Z}}
\newcommand{\Z}{\mathbb{Z}}

\renewcommand{\mod}{\text{ mod }}

\usepackage{datetime}
\usepackage{dsfont}

\usetikzlibrary{shapes.geometric}

\def\cX{{\mathcal X}}

\newcommand{\be}{\begin{equation}}
	\newcommand{\ee}{\end{equation}}
\newcommand{\bea}{\begin{eqnarray}}
	\newcommand{\eea}{\end{eqnarray}}

\newcommand{\doubleZ}{\mathbb{Z}}

\begin{document}

\begin{titlepage}
	
	\begin{flushright}
	\end{flushright}
	
	\vskip 3cm
	
	\begin{center}	
		{\Large \bfseries When are Duality Defects Group-Theoretical? }	
		
		\vskip 1cm
		Zhengdi Sun$^{1,2}$ and Yunqin Zheng$^{3,4,5,6}$
		\vskip 1cm
		
		\begin{centering}
			\begin{tabular}{ll}
				$^1$&Department of Physics, University of California, San Diego, CA 92093, USA, \\

                ${}^{2}$&Mani L. Bhaumik Institute for Theoretical Physics, Department of Physics and Astronomy, \\
                &University of California Los Angeles, CA 90095, USA,\\
				
				$^3$&Kavli Institute for the Physics and Mathematics of the Universe, \\
				& University of Tokyo,  Kashiwa, Chiba 277-8583, Japan\\
				
				$^4$&Institute for Solid State Physics, \\
				&University of Tokyo,  Kashiwa, Chiba 277-8581, Japan\\

                $^5$&C. N. Yang Institute for Theoretical Physics, \\&Stony Brook University, Stony Brook, NY 11794, USA\\

                $^6$&Kavli Institute for Theoretical Sciences, \\&University of Chinese Academy of Sciences, Beijing 100190, China
			\end{tabular}
		\end{centering}
		
		\vskip 1cm
		
	\end{center}
	
	\noindent
A quantum field theory with a finite abelian symmetry $G$ may be equipped with a non-invertible duality defect associated with gauging $G$. For certain $G$, duality defects admit an alternative construction where one starts with invertible symmetries with certain 't Hooft anomaly, and gauging a non-anomalous subgroup. This special type of duality defects are termed group theoretical. In this work, we determine when duality defects are group theoretical, among $G=\Z_N^{(0)}$ and $\Z_N^{(1)}$ in $2$d and 4d quantum field theories, respectively.  A duality defect is group theoretical if and only if its Symmetry TFT is a Dijkgraaf-Witten theory,
and we argue that this is equivalent to a certain stability condition of the topological boundary conditions of the $G$ gauge theory. By solving the stability condition, we find that a $\Z_N^{(0)}$ duality defect in 2d is group theoretical if and only if $N$ is a perfect square, and under certain assumptions a $\Z_N^{(1)}$ duality defect in 4d is group theoretical if and only if $N=L^2 M$ where $-1$ is a quadratic residue of $M$. For these subset of $N$, we construct explicit topological manipulations that map the non-invertible duality defects to invertible defects. We also comment on the connection between our results and the recent discussion of obstruction to duality-preserving gapped phases.

\end{titlepage}

\setcounter{tocdepth}{2}
\tableofcontents

\section{Introduction and summary}
\label{sec:intro}

\subsection{The problem}
\label{sec:intro1}

\paragraph{Duality defects:} 
A $d$-dimensional quantum field theory (QFT) $\CX$ with a finite, non-anomalous, $p$-form abelian global symmetry $G^{(p)}$ has a \emph{non-invertible duality symmetry} when $\CX$ is invariant under gauging $G^{(p)}$ \cite{Choi:2021kmx, Choi:2022zal}, 
\begin{eqnarray}\label{eq:selfdual}
    \CX= \CX/G^{(p)}.
\end{eqnarray}
This in particular requires $p=\frac{d}{2}-1$, so that the gauged theory $\CX/G^{(p)}$ also has the same $G^{(p)}$ global symmetry.\footnote{For simplicity, we only consider the symmetry of a single form-degree. One can also consider symmetries with multiple form-degrees, say, a $p$-form symmetry and a $q$-form symmetry. The theory can also be self-dual under gauging both symmetries if $d-2=p+q$. See e.g. \cite{Kaidi:2021xfk,Decoppet:2023bay} for such examples. More generally, a finite invertible symmetry in $d$ dimensional quantum field theory is described by a higher-group, and gauging it leads to the higher representation category of a higher group. Self-duality under gauging the higher group requires its higher representation category coincides with itself. See \cite{Bhardwaj:2022yxj,Bhardwaj:2022lsg,Freed:2022qnc, Bhardwaj:2022kot, Bartsch:2022ytj, Bhardwaj:2023wzd, Bhardwaj:2023ayw, Bartsch:2023wvv, Bartsch:2023pzl, Copetti:2023mcq, Bhardwaj:2022maz} for the recent developments of higher categorical theory description of global symmetries. } Each finite symmetry is associated with a topological defect \cite{Gaiotto:2014kfa}, and in the present case the \emph{duality defect}.  The duality defect can be constructed by gauging $G^{(p)}$ on half-space and imposing the Dirichlet boundary condition for $G^{(p)}$ at the defect locus. It has been shown that the duality defect satisfies \emph{non-invertible} fusion rule \cite{Choi:2021kmx,Choi:2022zal}\footnote{The non-invertible topological defects and some of their properties were explored during late 1980s and early 1990s, and see \cite{Alford:1992yx, Alford:1989ch, Bucher:1991bc, Alford:1990mk, Alford:1990ur, Alford:1990pt, Alford:1990fc} for a small selection of these works.}. By construction, duality defect implements gauging $G^{(p)}$. Note that there are additional data specifying the duality defects, such as the (symmetric and non-degenerate) bicharacter 
and the Frobenius-Schur indicator. Different QFT $\CX$ satisfying \eqref{eq:selfdual} may yield duality defects with different choices of such additional data. Throughout this work, for simplicity, we will denote the duality defect associated with gauging $G^{(p)}$ as \emph{$G^{(p)}$ duality defect}, to emphasize the underlying invertible symmetry, and leave the dependence of additional data implicit.

A generic $G^{(p)}$ symmetric QFT does not satisfy \eqref{eq:selfdual}. However, there are some well-known examples satisfying \eqref{eq:selfdual}, including compact scalars in 2d with $\Z_N^{(0)}$ symmetry \cite{Ji:2019ugf, Choi:2021kmx}, Maxwell theories in 4d with $\Z_N^{(1)}$ symmetry \cite{Choi:2021kmx}, $\CN=4$ super Yang-Mills theories in 4d with $\Z_N^{(1)}$ symmetry \cite{Kaidi:2021xfk}, etc. For these theories, showing \eqref{eq:selfdual} often requires a T-duality in 2d or S-duality in 4d,\footnote{One should distinguish the T or S-dualities from the duality transformation associated with gauging $G^{(p)}$.} and these dualities are usually found only in highly fine-tuned theories. Thus it is highly non-trivial and interesting to find deformations preserving the duality symmetry, i.e. the relation \eqref{eq:selfdual}. See \cite{Damia:2023ses} for recent discussions on duality-preserving deformation of the $\CN=4$ super Yang-Mills theory.

\paragraph{An alternative construction:} 
In \cite{Kaidi:2021xfk}, an alternative construction of theories with duality symmetry was proposed
for certain $G^{(p)}$. See also \cite{Bhardwaj:2022yxj, Bhardwaj:2022maz} for further generalizations.  The idea is to start with a theory $\CY$ with invertible symmetries only, and a mixed anomaly. Gauging a non-anomalous subgroup of $\CY$ yields another theory $\CX$, and the mixed anomaly in $\CY$ enforces the existence of duality symmetry in $\CX$. 

Let's illustrate the idea by an example. Take a 4d QFT $\CY$ with global symmetry $G^{(1)}\times H^{(0)}= \Z_2^{(1)}\times \Z_4^{(0)}$. We also assume a mixed anomaly characterized by the 5d anomaly theory\footnote{Only $\Z_2^{(0)}$ normal subgroup of $\Z_4^{(0)}$ is anomalous. But rigorously speaking, \eqref{eq:APB} is well-defined only when $A^{(1)}$ belongs to $H^1(X_5, \Z_4)$ (rather than $H^1(X_5, \Z_2)$) due to the $1/2$ factor.}
\begin{eqnarray}\label{eq:APB}
    \pi \int_{X_5} A^{(1)} \cup \frac{\CP(B^{(2)})}{2}.
\end{eqnarray}
The anomaly means that the partition function of $\CY$ obeys
\begin{eqnarray}\label{eq:Y}
    \CZ_{\CY}[B^{(2)}] = \CZ_{\CY}[B^{(2)}] e^{i\pi \int_{X_4} \CP(B^{(2)})/2}.
\end{eqnarray}
In particular, when $e^{i\pi \int_{X_4} \CP(B^{(2)})/2}$ evaluates to be a non-trivial phase, the partition function $\CZ_{\CY}[B^{(2)}]$ vanishes (for example due to the presence of zero modes). Next, we construct the theory $\CX$ by gauging $G^{(1)}=\Z_2^{(1)}$ of $\CY$ and then stacking an $\Z_2^{(1)}$ SPT,\footnote{We suppress the overall normalization throughout the paper. Please refer to \cite{Choi:2021kmx, Kaidi:2022cpf} for systematic discussions of the overall normalizations.} 
\begin{eqnarray}
    \CZ_{\CX}[B^{(2)}] = \sum_{b^{(2)}\in H^2(X_4, \Z_2)} \CZ_{\CY}[b^{(2)}] e^{i\pi \int_{X_4} b^{(2)} B^{(2)} + i \pi \int_{X_4} \CP(B^{(2)})/2}.
\end{eqnarray}
Combining with \eqref{eq:Y}, it is straightforward to check that $\CX=\CX/\Z_2^{(1)}$, i.e. 
\begin{eqnarray}
    \CZ_{\CX}[B^{(2)}] = \sum_{b^{(2)}\in H^2(X_4, \Z_2)} \CZ_{\CX}[b^{(2)}] e^{i\pi \int_{X_4} b^{(2)} B^{(2)}}.
\end{eqnarray}
The associated duality defect can be obtained using half-gauging. 
Hence we have found an alternative, yet systematic, way to construct theories with  duality symmetry associated with gauging $\Z_2^{(1)}$, as well as the $\Z_2^{(1)}$ duality defect.

It is useful to know that the duality symmetry associated with gauging $\Z_2^{(1)}$ admits the above alternative construction. Note that this construction does not require any detailed dynamical information of $\CY$ (such as whether $\CY$ is a CFT or a free field theory). Any $\CY$, as long as it has the requested symmetry and anomaly can be fed into the construction. In particular, because only the invertible symmetries of the theory $\CY$ play a role in the construction, it is easy to turn on perturbations leaving the symmetry and anomaly unchanged. After gauging $\Z_2^{(1)}$, such symmetric perturbation in $\CY$ becomes a duality-symmetry preserving perturbation in $\CX$. Hence it is easy to turn on duality-preserving deformations, and allows one to study the consequence of duality symmetry along the RG flow. Moreover, this alternative construction allows one to uncover new duality defects in gauge theories. For instance, this construction can be used to show the presence of duality defects in a large class of gauge theories, including non-invertible time reversal symmetries in the 4d Yang-Mills theories \cite{Kaidi:2021xfk, Bhardwaj:2022yxj, Choi:2022rfe}, non-invertible axial symmetries in 4d QED and QCD \cite{Choi:2022jqy,Cordova:2022ieu, vanBeest:2023dbu}, etc.

It turns out that for an arbitrary $G^{(p)}$, such an alternative construction may or may not exist. 
It is therefore useful to ask for which $G^{(p)}$ such an alternative construction exists. When it exists, the non-invertible duality defect in QFT $\CX$ can be mapped to an invertible defect in QFT $\CY$ under a \emph{topological manipulation} $\xi$, which includes gauging a non-anomalous subgroup and stacking an SPT etc, and such duality defect was named \emph{non-intrinsically non-invertible} \cite{Kaidi:2022uux}. Conversely, a duality defect which does not admit the alternative construction will be called intrinsically non-invertible.

On the other hand, for 2d QFTs with a generic finite Abelian symmetry $G^{(0)}$, the condition of when the alternative construction exists has been classified in mathematical literatures \cite{gelaki2009centers}. The duality defect was named \emph{group theoretical} if the alternative construction exists. In this work, we will follow the math notation and determine for which $G^{(p)}$ the duality defect is group theoretical.

As the answer in 2d is known, we will first review the results in \cite{gelaki2009centers}, and the goal is to present the discussion there using a more physical language, and pave the way for generalization to higher dimensions. For concreteness, we focus on the $\Z_N^{(0)}$ symmetry, although generalization to more complicated Abelian symmetries is possible.

We then generalize the 2d discussion and proceed to determine when duality defects associated with gauging $\Z_N^{(1)}$ in 4d QFTs are group theoretical. A partial list of group theoretical duality defects have been identified in \cite{Choi:2022zal, Bashmakov:2022uek}.  Our results not only reproduce the known ones in \cite{Choi:2022zal}, but also uncover new ones. As a systematic theory of higher category is less well-established than those in lower dimensions, we are unable to fully generalize the proofs in \cite{gelaki2009centers} to higher dimensions. Hence the completeness of our list of group theoretical duality defects only holds under certain assumptions which we will specify in Sec.~\ref{sec:criterias}.

\subsection{Main results}
\label{sec:intro2}

The key idea to find the group theoretical duality defect is to use the \emph{Symmetry Topological Field Theory} (SymTFT) \cite{Kaidi:2022cpf, Apruzzi:2021nmk, Freed:2022qnc, Kong:2020cie,Ji:2019jhk, Gaiotto:2020iye, Burbano:2021loy, Bhardwaj:2023ayw, Lin:2022dhv}.\footnote{See also \cite{Inamura:2023qzl, Bhardwaj:2023ayw, Bartsch:2023wvv, Lin:2023uvm, Kaidi:2023maf, Zhang:2023wlu, Moradi:2022lqp, vanBeest:2022fss, Chen:2023qnv, Apruzzi:2023uma, Apruzzi:2022rei, Apruzzi:2022dlm} for applications of the SymTFT to the dynamics of QFTs, lattice models and string/M-theories. } A $d$-dimensional QFT with a global symmetry described by a (higher) fusion category $\CC$ is equivalent to a $(d+1)$- dimensional ``sandwich" where in the bulk is a SymTFT describing the Drinfeld center of $\CC$, the left $d$-dimensional boundary condition is a canonical/Dirichlet boundary condition where the defects labeled by $\CC$ are supported, and the right boundary condition is a non-topological boundary condition encoding all the dynamical information of the QFT. Since the bulk is topological, one can shrink the sandwich by colliding the left and right boundaries to recover the $d$-dimensional QFT.

One of the advantages of the sandwich construction is that it automatically encodes the additional data of the duality defect mentioned in Sec.~\ref{sec:intro1}, i.e. the bicharacters and the Frobenius-Schur indicator, while they are not explicit from \eqref{eq:selfdual}.

The sandwich construction also enjoys two interesting properties \cite{Kaidi:2022cpf, Apruzzi:2021nmk, Freed:2022qnc}. First, when the symmetry of the QFT is invertible, the SymTFT is a gauged anomaly theory, i.e. a Dijkgraaf-Witten theory, whose partition function is
\begin{eqnarray}\label{eq:DWdef}
    \CZ= \sum_{g\in G} e^{i \omega(g)}
\end{eqnarray}
where $G$ is the finite (higher) gauge group of the Dijkgraaf-Witten theory, $g$ is a (higher form) gauge field valued in $G$,\footnote{For simplicity we use $g\in G$ to represent a $g$ gauge field taking value in $G$, although a more appropriate way is to write $g\in H^*(X_{d+1}, G)$. We hope this simplified notation does not lead to confusion.} and $\omega(g)$ is the twist term which specifies the $G$-anomaly.\footnote{There are more restricted definition of the Dijkgraaf-Witten in the literature, where $G$ is an ordinary group. For more general $G$, e.g. $G$ is a 2-group, the corresponding 4d TFT is called Yetter TFT \cite{Yetter:1993dh}. See \cite{Inamura:2023qzl} for a review of various TFTs in higher dimensions. Here we denote the $G$ gauge theory with any finite (higher) group $G$ as Dijkgraaf-Witten. }
Second, because all the symmetry defects (or background fields when the symmetry is invertible) are supported on the left topological boundary, topological manipulations (including gauging and stacking an invertible phase) do not change the SymTFT. Combining these two properties, we conclude that \emph{a duality defect is group theoretical if and only if its SymTFT is a Dijkgraaf-Witten theory}.

The SymTFT of a duality defect associated with gauging $G^{(p)}$ admits a convenient construction \cite{Kaidi:2022cpf, Zhang:2023wlu}. One starts with the SymTFT of $G^{(p)}$ symmetry (without a duality defect), which is simply a $G^{(p)}$ $(p+1)$-form gauge theory in $2p+3$ dimensions. Such a theory admits an electro-magnetic (EM) exchange symmetry whose order depends the parity of $p$. The SymTFT of the duality defect (including $G^{(p)}$ symmetry) is obtained by gauging the EM exchange symmetry of the $G^{(p)}$ $(p+1)$-form gauge theory. As we will review in Sec.~\ref{sec:criterias}, such SymTFT is a Dijkgraaf-Witten theory amounts to the existence of an EM stable topological boundary condition of the $G^{(p)}$ $(p+1)$-form gauge theory. This latter condition will be explicitly checked in the following sections.

In this paper, we focus on the duality defects associated with gauging $G^{(p)}= \Z_N^{(0)}$ in 2d, and $G^{(p)}= \Z_N^{(1)}$ in 4d. We determine for which $N$ together with the choice of bicharacters and the Frobenius-Schur indicator, the duality defect is group theoretical by examining when the EM stable topological boundary condition of $\Z_N^{(0)}$ (or $\Z_N^{(1)}$) gauge theory exists in 3d (or 5d). We find the following results:
\paragraph{$\Z_N^{(0)}$ duality defects in 2d:} The $\Z_N^{(0)}$ duality defect is group theoretical if and only if $N$ is a perfect square. 
\paragraph{$\Z_N^{(1)}$ duality defects in 4d:} On spin manifolds, the $\Z_N^{(1)}$ duality defect is group theoretical if and only if $N=L^2 M$ where $-1$ is a quadratic residue of $M$. \\~\\ 
The above conclusion holds for arbitrary choices of additional data, including the bicharacters and the Frobenius-Schur indicators.

The 2d results were already proven in a mathematics reference \cite{gelaki2009centers}. In the main text, for both 2d and 4d cases, we will show the if direction by explicitly demonstrating the SymTFT to be a Dijkgraaf-Witten, and also working out the explicit topological manipulation. In particular, the special case of $L=1$ in 4d was known in \cite{Choi:2022zal}, and our result shows that there are new cases for $L>1$. However, unlike the 2d case where the only if direction is proven,\footnote{In \cite{Kaidi:2022cpf}, the authors used the SymTFT to give a physical derivation of the only if direction in 2d. The observation there is that the SymTFT of the duality defect has line operators of quantum dimension $\sqrt{N}$. Note that all line operators in any bosonic Dijkgraaf-Witten theory are of integer quantum dimension \cite{Hu:2012wx, Dijkgraaf:1989pz, deWildPropitius:1995cf}, the SymTFT can be Dijkgraaf-Witten only if $N$ is a perfect square. } for the 4d case the only if direction remains a conjecture.

We also note that the question of whether a duality defect is group theoretical has also been extensively discussed in the context of string/M-theories. See \cite{Heckman:2022xgu, Lawrie:2023tdz, Apruzzi:2023uma}. In particular, in \cite{Apruzzi:2023uma}  whether the $\Z_N^{(1)}$ duality defect is group theoretical was phrased in terms of Hanany-Witten transition between strings and 7-branes in the holographic IIB setup, and the authors found the same sequence of $N$ for $N\leq 29$.\footnote{After our preprint appeared on arXiv, we were informed by Fabio Apruzzi that \cite{Apruzzi:2023uma} contains overlapping results with the present work.}\footnote{In \cite{Cordova:2023bja,Antinucci:2023ezl} the anomaly of duality defects is derived, and the condition for duality defects being group theoretical is found to be a necessary, but non sufficient, condition for them to be anomaly free. }

Note that we assumed the 4d spacetime to be a spin manifold. On non-spin manifolds, the criteria for odd $N$ remains the same. However, for even $N$, the situation is more complicated, and we will comment on them in the main text. 

The organization of the paper is as follows. In Sec.~\ref{sec:criterias}, we discuss the general strategy to determine when a duality defect is group theoretical. This section is largely based on \cite{gelaki2009centers}, presented in a way that applies to higher dimensions as well. Sec.~\ref{sec:1+1} and Sec.~\ref{sec:3+1} are in parallel, which discuss the group theoretical duality defects in 2d and 4d respectively. In both sections, we first identify the group theoretical $N$'s using the stable topological boundary condition following the strategy of Sec.~\ref{sec:criterias}. Then for each $N$, we explicitly demonstrate that the SymTFT is a Dijkgraaf-Witten theory, and also propose the explicit topological manipulation which maps the duality defect to the invertible defect. In Sec.~\ref{sec:comments}, we comment on the relation between group theoretical duality defects and obstructions to duality preserving gapped phases.

\section{Criteria of group-theoretical duality defects}
\label{sec:criterias}
In this section, we study in general when the duality defects are group theoretical. A rigorous mathematical discussion of this problem in 2d was already available in \cite{gelaki2009centers}. The goal of this section is to translate the discussion to a more physicist-friendly language. We also try to present the discussion in a way applicable to 4d.

\subsection{Symmetry TFT of duality defects}
\label{sec:symtftintro}

As reviewed in the introduction, the SymTFT is a useful tool to identify whether a fusion category is group theoretical. We thus first review the properties of the SymTFT of duality defects, focusing on $\Z_N^{(d/2-1)}$ symmetry in $d$ dimensions, for $d=2,4$. We will follow the discussion in  \cite{Kaidi:2022cpf}.

Consider a $d$ dimensional QFT $\CX$ with a non-anomalous $\Z_N^{(d/2-1)}$  symmetry. Let's denote its partition function as $\CZ_{\CX}[B^{(d/2)}]$. Any such theory can be expanded into a $d+1$ dimensional slab, as shown in Fig.~\ref{fig:symmTFTidea}. In the bulk of the slab, there is a $d+1$ dimensional $\Z_N$ $d/2$-form gauge theory, whose Lagrangian is
\begin{eqnarray}\label{eq:leftbdy}
    \CL= \frac{2\pi}{N} \widehat{b}^{(d/2)} \delta b^{(d/2)}.
\end{eqnarray}
This is the SymTFT of the $\Z_N^{(d/2-1)}$ symmetry. 
On the left boundary, there is a Dirichlet topological boundary condition/state obtained by setting the electric field $b^{(d/2)}$ to background value, 
\begin{eqnarray}\label{eq:rightbdy}
    \bra{\text{Dir}(B^{(d/2)})} = \sum_{b^{(d/2)}\in H^{d/2}(X_d, \Z_N)}  \delta(b^{(d/2)}- B^{(d/2)})\bra{b^{(d/2)}}.
\end{eqnarray}
On the right boundary, there is a dynamical boundary encoding all the information of the QFT $\CX$, where the boundary state is 
\begin{eqnarray}\label{eq:dynbdystate}
    \ket{\CX}= \sum_{b^{(d/2)}\in H^{d/2}(X_d, \Z_N)} \CZ_{\CX}[b^{(d/2)}] \ket{b^{(d/2)}}.
\end{eqnarray}
Shrinking the slab amounts to colliding the boundary states \eqref{eq:leftbdy} and \eqref{eq:rightbdy}, which reproduces the partition function $\CZ_{\CX}[B^{(d/2)}]$.

\begin{figure}[!tbp]
	\centering
	\begin{tikzpicture}[scale=0.8]

	\shade[line width=2pt, top color=blue!30, bottom color=blue!5] 
	(0,0) to [out=90, in=-90]  (0,3)
	to [out=0,in=180] (6,3)
	to [out = -90, in =90] (6,0)
	to [out=180, in =0]  (0,0);
	
		\draw[very thick] (-7,0) -- (-7,3);
	\node[below] at (-7,0) {$\CZ_{\CX}[B^{(d/2)}]$};

	\draw[thick, snake it, <->] (-1.7,1.5) -- (-5, 1.5);
	
	\draw[thick] (0,0) -- (0,3);
	\draw[thick] (6,0) -- (6,3);
	\node at (3,1.5) {\begin{tabular}{c}
	     $\Z_N$ $d/2$-form  \\
	     gauge theory
	\end{tabular} };
	\node[below] at (0,0) {$\langle \text{Dir}(B^{(d/2)})|$};
	\node[below] at (6,0) {$|\cX\rangle $}; 
	
	\end{tikzpicture}

	\caption{A $d$ dimensional QFT $\CX$ with a non-anomalous $\Z_N^{(d/2-1)}$ symmetry can be expanded into a $d+1$ dimensional slab, where the bulk of the slab is the SymTFT of the $\Z_N^{(d/2-1)}$ symmetry---$\Z_N$ $d/2$-form gauge theory, the left boundary is a topological boundary condition setting the electric 2-form gauge field to classical background value $B^{(2)}$, and the right boundary is a non-topological dynamical boundary condition. }
	\label{fig:symmTFTidea}
\end{figure}

We further require the QFT $\CX$ to be invariant under gauging $\Z_N^{(d/2-1)}$, i.e. 
$\CX=\CX/\Z_N^{(d/2-1)}$, so that the symmetry of $\CX$ contains not only $\Z_N^{(d/2-1)}$ but also self-duality. To obtain the SymTFT of the full symmetry, we start with the $\Z_N$ $d/2$-form gauge theory in $d+1$ dimensions and gauge the electro-magnetic (EM) exchange symmetry. This symmetry basically exchanges $b^{(d/2)}$ and $\widehat{b}^{(d/2)}$, to be more precise,
\begin{eqnarray}\label{eq:emexchangesym}
    b^{(d/2)} \to u \widehat{b}^{(d/2)}, \qquad \widehat{b}^{(d/2)} \to (-1)^{d/2+1} v b^{(d/2)},
\end{eqnarray}
where $uv=1$ mod $N$. By Chinese Remainder Theorem, both $u$ and $v$ are coprime with $N$. 
The minus sign means that the EM exchange symmetry is $\Z_2^{\text{em}}$ when $d=2$ mod $4$, and is $\Z_4^{\text{em}}$ when $d=0$ mod $4$.\footnote{When $N=2$ and $d=0$ mod $4$, the EM exchange symmetry is still $\Z_2^{(0)}$ because $b^{(d/2)}= - b^{(d/2)}$ mod $2$.} Below for simplicity the EM symmetry will be denoted as $\Z_{\chi}^{\text{em}}$, with $\chi= 4/2^{[d/2]_2}$.  
Note that when gauging  $\Z_{\chi}^{\text{em}}$, there is a freedom of choosing the discrete theta term labeled by $\epsilon\in H^{d+1}(\Z_{\chi}^{\text{em}}, U(1))$. When $d=2$, different choices of $u$ (and hence $v$) correspond to bicharacters of the Tambara-Yamagami fusion category $\text{TY}(\Z_N^{(0)},u,\epsilon)$, while different choices of the discrete theta term $\epsilon$ correspond to the Frobenius-Schur indicator \cite{TAMBARA1998692, Zhang:2023wlu,Thorngren:2019iar}. For simplicity, we will adopt the same notation in $d=4$ as well. 
In short, 
\begin{eqnarray}\label{eq:symtft}
    \text{SymTFT of duality symmetry } = \text{ $\Z_N$ $d/2$-form gauge theory}/(\Z_{\chi}^{\text{em}})_{u,\epsilon}.
\end{eqnarray}

\subsection{Criteria for group-theoretical duality defects}

We proceed to use the SymTFT to determine when the $\Z_N^{(d/2-1)}$ duality defect is group theoretical. 
The schematic idea is shown in Fig.~\ref{fig:idea}. We first recall, as discussed in the introduction, that a duality defect is group theoretical if and only if its SymTFT is a Dijkgraaf-Witten theory. Using \eqref{eq:symtft}, the problem boils down to showing that the $\Z_N$ $d/2$-form gauge theory $/ (\Z_\chi^{\text{em}})_{u,\epsilon}$ is a Dijkgraaf-Witten theory. This completes the first two arrows on the top of Fig.~\ref{fig:idea}.

\begin{figure}
    \centering
    \begin{tikzpicture}
    [auto,
 block/.style ={rectangle, draw=black, thick, fill=blue!20, text width=25em,align=center, rounded corners, minimum height=2em},
 block1/.style ={rectangle, draw=blue, thick, fill=blue!20, text width=5em,align=center, rounded corners, minimum height=2em},
 line/.style ={draw, thick, -latex',shorten >=2pt},
 cloud/.style ={draw=red, thick, ellipse,fill=red!20,
 minimum height=1em}]
        \node[block] at (0,0) {$\Z_N^{(d/2-1)}$ duality defect is group theoretical};
        \draw[thick, ->] (-0.2, -0.5) -- (-0.2, -1.5);
        \draw[thick, <-] (0.2, -0.5) -- (0.2, -1.5);
        \node[block] at (0, -2) {$\Z_N$ $d/2$-form gauge theory$/(\Z_{\chi}^{\text{em}})_{u,\epsilon}$ is Dijkgraaf-Witten};
        \draw[thick, ->] (-0.2, -2.5) -- (-0.2, -3.5);
        \draw[thick, <-] (0.2, -2.5) -- (0.2, -3.5);
        \node[block] at (0, -4.3) {$\exists$ Lagrangian subcategory $\CS$ in $\Z_N$ $d/2$-form gauge theory$/(\Z_{\chi}^{\text{em}})_{u,\epsilon}$};
        \draw[thick, ->, dashed] (-0.2, -5) -- (-0.2, -6.3);
        \draw[thick, <-] (0.2, -5) -- (0.2, -6.3);
        \node[block] at (0, -7) {$\exists \Z_{\chi}^{\text{em}}$ stable Lagrangian subgroup $\CS_{\text{pre}}$ in $\Z_N$ $d/2$-form gauge theory};
    \end{tikzpicture}
    \caption{Idea of determining when a $\Z_N^{(d/2-1)}$ duality defect is group theoretical. We will argue (at the physics level of rigor) for the solid arrows, while for the dashed arrow, our argument is based on an assumption that is only proved in $d=2$.}
    \label{fig:idea}
\end{figure}

Below, we argue for the remaining arrows.  Sec.~\ref{sec:DW=Lagsubcategory} establishes the two arrows in the middle of Fig.~\ref{fig:idea}. Sec.~\ref{sec:sufcond} and Sec.~\ref{sec:nececond} discusses the upward and dashed downward arrow at the bottom, respectively.

\subsubsection{Dijkgraaf-Witten $=$ existence of Lagrangian subcategory}
\label{sec:DW=Lagsubcategory}

Let's start with the observation that a SymTFT \eqref{eq:symtft} is Dijkgraaf-Witten if and only if there exist a set of topological operators $S_\alpha$'s such that 
\begin{enumerate}
    \item $S_\alpha$'s form a (higher) representation category of some symmetry group $G$, and for simplicity we schematically denote them as $\Rep(G)$. \footnote{The higher representation category, e.g. $2\Rep(G)$, has been discussed recently in the context of generalized symmetries and generalized charges in 
 \cite{Bartsch:2022mpm, Bhardwaj:2022lsg, Bartsch:2023pzl, Bhardwaj:2023wzd, Bartsch:2022ytj, Bhardwaj:2022maz, Bhardwaj:2023ayw}. In this section, we will schematically denote both the ordinary and higher representation categories as $\Rep(G)$. } 
    \item $S_\alpha$'s are gaugable and gauging them leads to an invertible theory. \footnote{Gauging $\CS$ amounts to first form a algebra object $\CA$ from $S_{\alpha}$'s, and insert a mesh of $\CA$ in the path integral. For 3d TFT, an algebra such that after gauging it leads to an invertible theory is known as Lagrangian algebra, which is shown to be a gaugable algebra, whose quantum dimension is the total quantum dimension of the TFT. In higher dimension, the theory of Lagrangian algebra is less well established, and we will simply define the Lagrangian algebra to be gaugable (i.e. can consistently insert a mesh of it) and has the property that gauging it would lead to an invertible theory. See \cite{Zhao:2022yaw} for a recent discussion of the Lagrangian algebra of 4d TFTs. Mathematically rigorous discussions on gauging a fusion 2-category can be found in \cite{Decoppet:2022dnz, Decoppet:2023bay}.
    }
\end{enumerate}
Let's denote the subcategory whose objects are $S_\alpha$ as the \emph{Lagrangian subcategory} $\CS$. One direction of this claim is obvious: in the Dijkgraaf-Witten theory with gauge group $G$ and cocycle $\omega$, obviously the set of Wilson operators form $\Rep(G)$. Hence the first property is satisfied. Furthermore, $\Rep(G)$ is gaugable because it is the quantum symmetry from gauging $G$ of the $G$ symmetric invertible theory (i.e. $G$-SPT) $\omega$.  This further means that gauging $\Rep(G)$ leads to an invertible theory. This shows the second property, and completes the only if direction, i.e. the $\downarrow$ in the middle of Fig.~\ref{fig:idea}.

For the if direction, suppose a SymTFT has a set of gaugable topological operators labelled by $\Rep(G)$ and gauging them leads to an invertible theory whose partition function is a phase $e^{i \omega}$, then the quantum symmetry is $G$ and we can gauge $G$ to recover the original SymTFT. This ensures the original SymTFT is Dijkgraaf-Witten. This completes the if direction, i.e. the $\uparrow$ in the middle of Fig.~\ref{fig:idea}.

The above definition of the ``Lagrangian subcategory" was motivated by the results of 3d TFT.   
In the context of 3d TFT, the set of anyons satisfying the above conditions form a fusion subcategory known as the Lagrangian subcategory $L$ \cite{gelaki2009centers,2007arXiv0704.0195D}. It satisfies the following two conditions
\begin{enumerate}
    \item $L$ is of the form $\Rep(G)$ equipped with the standard symmetric braiding for some finite group $G$. The subcategory of this form is called \emph{Tannakian}.
    \item $L = L'$ where $L'$ is the centralizer defined as the fusion subcategory (of the entire braided modular tensor category of the 3d TFT) which contains all the anyon $a$ having trivial braiding with every anyon in $L$. 
\end{enumerate}
Notice that the first condition ensures the 1-form symmetries generated by $\Rep(G)$ is gaugable, and to gauge it we can consider condensing the algebra $A_G$ which is the regular representation in $\Rep(G)$. The second condition implies every line operator outside $L$ is charged non-trivially under the 1-form symmetry generated by the lines in $L$, hence condensing $L$ would project out every line operator and end up with an invertible theory. In other words, the algebra $A_G$ is Lagrangian. We reformulate the second condition such that the statement works for higher dimension as well. For this reason, we will also name the higher dimensional generalization of $L$, i.e. $\mathcal{S}$, as the Lagrangian subcategory.

Given a Lagrangian algebra $A_G$ as the regular representation in $\Rep(G)$, we can consider half-gauging it to engineer a gapped boundary for the SymTFT. In the half space with invertible theory, there's apparently a quantum $G$ symmetry whose symmetry defects can terminate on the gapped boundary and the intersections are the $G$-symmetry defects on the gapped boundary.

\subsubsection{Up arrow at the bottom of Fig.~\ref{fig:idea}}
\label{sec:sufcond}

We proceed to argue for the up arrow at the bottom of Fig.~\ref{fig:idea}, i.e.
\begin{eqnarray}\label{eq:ifcondition}
    \begin{tikzpicture}
    [auto,
 block/.style ={rectangle, draw=black, thick, fill=blue!20, text width=25em,align=center, rounded corners, minimum height=2em},
 block1/.style ={rectangle, draw=blue, thick, fill=blue!20, text width=5em,align=center, rounded corners, minimum height=2em},
 line/.style ={draw, thick, -latex',shorten >=2pt},
 cloud/.style ={draw=red, thick, ellipse,fill=red!20,
 minimum height=1em}]
        \node[block] at (0, -4.3) {$\exists$ Lagrangian subcategory $\CS$ in $\Z_N$ $d/2$-form gauge theory$/(\Z_{\chi}^{\text{em}})_{u,\epsilon}$};
        \draw[thick, <-] (0.2, -5) -- (0.2, -6.3);
        \node[block] at (0, -7) {$\exists \Z_{\chi}^{\text{em}}$ stable Lagrangian subgroup $\CS_{\text{pre}}$ in $\Z_N$ $d/2$-form gauge theory};
    \end{tikzpicture}
\end{eqnarray}
Here we denote the Lagrangian subgroup of the $\Z_N$ $d/2$-form gauge theory (without gauging $\Z_{\chi}^{\text{\text{em}}}$) as $\CS_{\text{pre}}$, where the subscript pre is to distinguish it from the Lagrangian subcategory $\CS$ of the $\Z_\chi^{\text{em}}$ gauged theory \eqref{eq:symtft}. Since $\Z_N$ $d/2$-form gauge theory is an abelian TQFT, its boundary condition is specified by the Lagrangian (higher) subgroup \cite{Chen:2021xuc}. The operators in the Lagrangian subgroup automatically form a $d/2$ representation category, i.e. $\CS_{\text{pre}}= \Rep(G)$, for certain finite group $G$.  Thus the Lagrangian subgroup is automatically a Lagrangian subcategory. For this reason, we will use the Lagrangian subgroup to denote the Lagrangian subcategory of an Abelian TQFT throughout this paper.

A Lagrangian subgroup $\CS_{\text{pre}}$ is $\Z_{\chi}^{\text{\text{em}}}$ stable means that $\Z_{\chi}^{\text{\text{em}}}$ is a group endomorphism of $\CS_{\text{pre}}$, i.e. under $\Z_{\chi}^{\text{\text{em}}}$, $\CS_{\text{pre}}$ is mapped to $\CS_{\text{pre}}$ itself, while each object in $\CS_{\text{pre}}$ might transform non-trivially.  Suppose $\CS_{\text{pre}}$ is $\Z_{\chi}^{\text{\text{em}}}$ stable, 
we can recover a $\doubleZ_\chi^\rho$ automorphism acting the group $G$ itself from this. In the $\Z_{\chi}^{\text{\text{em}}}$ gauged theory, we can then construct a fusion subcategory of the form $\Rep(G\rtimes_\rho \doubleZ_\chi)$, generated by $\doubleZ_\chi$ orbit of $\Rep(G)$ and the quantum $\doubleZ_\chi$ symmetry defect $K$. Indeed, an irreducible representation in $G\rtimes_\rho \doubleZ_\chi$ is either constructed as a direct sum of two irreducible representations of $G$ related by $\doubleZ_\chi$, or a tensor product between a $\doubleZ_\chi$-invariant irreducible representation of $G$ and a representation of the $\doubleZ_\chi^\rho$.

It remains to show that $\CS' = \Rep(G\rtimes_\rho \doubleZ_\chi)$ is gaugable and that gauging it leads to an invertible theory. This is the case because gauging the entire $\Rep(G\rtimes_\rho \doubleZ_\chi)$ can be achieved by sequential gauging where we first gauge $\doubleZ_\chi^{(d-2)}$ generated by the quantum $\Z_\chi$ defect $K$ to recover $\Rep(G)$ and then gauge the anomaly free $\Rep(G)$ which will leads to the trivial theory. Hence $\Rep(G\rtimes_\rho \doubleZ_\chi)$ is the Lagrangian subcategory of the $\Z_\chi^{\text{em}}$ gauged theory.  This completes the proof.

We comment that the $\Z_\chi^{\text{\text{em}}}$ stable Lagrangian subgroup is equivalent to $\Z_\chi^{\text{\text{em}}}$ stable boundary condition or state of the $\Z_N$ $d/2$-form gauge theory. Given a Lagrangian subgroup consisting of the operators $S_\alpha$, where $\alpha\in \CI$ and $\CI$ is the index set  labeling the simple object in $\CS_{\text{pre}}$, the topological boundary state $\ket{\psi_{\CI}}$ is determined via $S_{\alpha}\ket\psi=\ket\psi$ for all $\alpha\in \CI$. Note that the boundary state is unique, and can be constructed by gauging the Lagrangian subgroup on half space. Denote the symmetry operator of $\Z_{\chi}^{\text{\text{em}}}$ as $U$, stable Lagrangian subcategory means $U S_{\alpha} U^{-1} = S_{\beta}$ where $\beta\in \CI$. It follows that
\begin{eqnarray}
    S_{\beta} U\ket\psi= U S_{\alpha} U^{-1} U \ket\psi = US_{\alpha}\ket\psi = U\ket\psi
\end{eqnarray}
meaning that $U\ket\psi$ is stabilized by the same Lagrangian subgroup. By uniqueness, we have $U\ket\psi = \ket\psi$, which means the boundary state $\ket\psi$ is also $\Z_{\chi}^{\text{\text{em}}}$ stable. 

In Sec.~\ref{sec:1+1} and Sec.~\ref{sec:3+1}, we will enumerate all the Lagrangian subgroups of the $\Z_N$ $d/2$-form gauge theory and classify when they are $\Z_\chi^{\text{\text{em}}}$ stable. When there exist stable Lagrangian subgroups, we will explicitly show that gauging the $\Z_\chi^{\text{\text{em}}}$ symmetry leads to Dijkgraaf-Witten theory, and also find explicit topological manipulations that map the duality defect to an invertible defect. Such an explicit construction gives a practical proof of the up arrows, from the bottom to the top, in Fig.~\ref{fig:idea}.

\subsubsection{Dashed down arrow at the bottom of Fig.~\ref{fig:idea}}
\label{sec:nececond}

We proceed to show the dashed down arrow at the bottom of Fig.~\ref{fig:idea}, i.e.  
\begin{eqnarray}\label{eq:onlyifcondition}
    \begin{tikzpicture}
    [auto,
 block/.style ={rectangle, draw=black, thick, fill=blue!20, text width=25em,align=center, rounded corners, minimum height=2em},
 block1/.style ={rectangle, draw=blue, thick, fill=blue!20, text width=5em,align=center, rounded corners, minimum height=2em},
 line/.style ={draw, thick, -latex',shorten >=2pt},
 cloud/.style ={draw=red, thick, ellipse,fill=red!20,
 minimum height=1em}]
        \node[block] at (0, -4.3) {$\exists$ Lagrangian subcategory $\CS$ in $\Z_N$ $d/2$-form gauge theory$/(\Z_{\chi}^{\text{em}})_{u,\epsilon}$};
         \draw[thick, ->, dashed] (-0.2, -5) -- (-0.2, -6.3);
        \node[block] at (0, -7) {$\exists \Z_{\chi}^{\text{em}}$ stable Lagrangian subgroup $\CS_{\text{pre}}$ in $\Z_N$ $d/2$-form gauge theory};
    \end{tikzpicture}
\end{eqnarray}
We use the dashed arrow to emphasize that it is proven only in $d=2$ \cite{gelaki2009centers}, while 
for higher $d$, we are only able to generalize the results in \cite{gelaki2009centers} under an assumption which we specify below.

Suppose the $\Z_N$ $d/2$-form gauge theory $/(\Z_{\chi}^{\text{em}})_{u,v,\epsilon}$ contains a Lagrangian subcategory $\CS= \Rep(G)$. Since the theory comes from gauging $\doubleZ_\chi^{\text{\text{em}}}$, there must be an $\doubleZ_\chi^{\text{\text{em}}}$ bosonic topological line operator $K$ corresponding to the quantum $\doubleZ_\chi$ $(d-1)$-form symmetry of the gauged theory. Clearly, the subcategory generated by $K$ is of the form $\Rep(\doubleZ_\chi^K)$.  Below we will make a key assumption: 
\paragraph{Key assumption:}
$\Rep(\doubleZ_\chi^K)$ is a subcategory of $\CS=\Rep(G)$.\\~\\
Since $K$ obeys the $\doubleZ_\chi$ fusion rule, it must be labelled by an order $\chi$ representation of $G$. Equivalently, we can describe $K$ as a surjective homomorphism $G\rightarrow \doubleZ_\chi$, and denote the kernel of this homomorphism as $H$ then we have the following short exact sequence:
\begin{equation}\label{eq:extensionH}
    0 \rightarrow H \rightarrow G \rightarrow \doubleZ_\chi \rightarrow 0.
\end{equation}
After gauging $\doubleZ_\chi^{K}$ by condensing the $K$-line, the Lagrangian subcategory $\Rep(G)$ then becomes $\Rep(H)$. To see $\Rep(H)$ is Lagrangian, notice that gauging $H$ is the second step in the sequential gauging, which corresponds to gauge $\Rep(G)$ in the original theory. Therefore, gauging $\Rep(H)$ must lead to an invertible theory and $\Rep(H)$ is Lagrangian. It is also clear that $\Rep(H)$ must be stable under the $0$-form $\doubleZ_\chi^{\text{\text{em}}}$ symmetry from condensing $K$. Hence, up to the assumption that the Lagrangian subcategory $\Rep(G)$ contains $\Rep(\Z_\chi^K)$ as a subcategory, we argued that \eqref{eq:onlyifcondition} holds.

When $d=2$, however, there is a theorem guarantees that the \textbf{key assumption} is true: we can always find a $\Rep(G)$ such that it contains the $K$-line\cite{drinfeld2010braided,etingof2016tensor}. Given a Tannakian subcategory  $\Rep(H)$,\footnote{Note that $H$ here is a generic group and should not be confused with the $H$ in \eqref{eq:extensionH}.} in general it may be contained as a subcategory in another Tannakian subcategory $\Rep(H')$. A Tannakian subcategory which is not properly\footnote{Given two categories $\CC$ and $\CD$, $\CD$ is properly contained in $\CC$ if $\CD$ is a subcategory of $\CC$ but $\CD$ is not the same as $\CC$.} contained in another Tannakian subcategory is called maximal \cite{drinfeld2010braided,etingof2016tensor}. Notice that a Lagrangian subcategory is automatically a maximal Tannakian subcategory. Since a Tannakian subcategory describes gaugable 1-form symmetries, one can consider condense those anyons to get a smaller TFT. In \cite{drinfeld2010braided,etingof2016tensor}, it was shown that condensing maximal Tannakian subcategory would lead to equivalent TFT which doesn't depend on the choice of the maximal Tannakian subcategory.\footnote{The TFT acquired from the gauging is called the core of the original TFT, which seems to give a measure on the intrinsically non-invertibleness of a symmetry TFT.} Using this, we can start with the Tannakian subcategory $\Rep(\doubleZ_2^K)$ and find the maximal Tannakian subcategory $\Rep(G)$ containing it. Then, by the above theorem, condensing $\Rep(G)$ would lead to trivial theory and therefore $\Rep(G)$ is Lagrangian. This completes the proof for the necessary condition for $d=2$. For higher dimensions, we do not know if the analog of such theorem  exists, so \eqref{eq:onlyifcondition} is only a conjecture.

\section{Group theoretical duality defects in 2d}
\label{sec:1+1}

The SymTFT of $\Z_N^{(0)}$ duality defects is a 3d $\Z_N$ gauge theory with $\Z_2^{\text{\text{em}}}$ symmetry gauged. It has been shown in \cite{gelaki2009centers}, which we revisited in Sec.~\ref{sec:criterias}, that the duality defect is group theoretical if and only if the 3d $\Z_N$ gauge theory admits $\Z_2^{\text{\text{em}}}$ stable topological boundary condition. In this section, we first review the topological boundary condition of the 3d $\Z_N$ gauge theories, and determine when a $\Z_2^{\text{\text{em}}}$ stable topological boundary condition is allowed. We then explicitly show that for those allowed $N$ the SymTFT is a Dijkgraaf-Witten theory, and find explicit topological manipulations under which the duality defects are mapped to invertible defects.

\subsection{Lagrangian subgroups of 3d $\Z_N$ gauge theory}
\label{sec:tbc3d}

The action of the $\Z_N$ gauge theory is 
\begin{eqnarray}\label{eq:ZNaction}
	\CL = \frac{2\pi}{N} \widehat{a} \delta a
\end{eqnarray} 
where $\widehat{a}, a$ are both $\Z_N$ cochains. It has a $\Z_2^{\text{\text{em}}}$ exchange symmetry
\begin{eqnarray}\label{eq:emsymmetry}
    a\to  u \widehat{a}, \qquad \widehat{a}\to  v a,
\end{eqnarray}
with $uv=1$ mod $N$. 
The topological lines are
\begin{eqnarray}
    L_{(e,m)}(\gamma)= e^{\frac{2\pi i e}{N}\oint_{\gamma} a} e^{\frac{2\pi i m}{N}\oint_{\gamma} \widehat{a}}.
\end{eqnarray}

The topological boundary conditions of an Abelian TQFT are classified by the Lagrangian subgroups. The Lagrangian subgroup $\CA$ consists of $N$ topological line operators $L_{(e,m)}$ with the following conditions,
\begin{enumerate}
    \item $L_{(e,m)}$ has trivial topological spin, i.e. $e^{i 2\pi e m/N}=1$. 
    \item Any two lines $L_{(e,m)}$ and $L_{(e',m')}$ in the Lagrangian subgroup $\CA$ have trivial mutual braiding, i.e. $e^{2\pi i (em'+ e'm)/N}=1$. 
    \item Any other line operator that does not belong to $\CA$ braids non-trivially with at least one line in $\CA$. 
\end{enumerate}
We remark that the first condition automatically implies the second condition, since the mutual braiding phase $B_{(e,m),(e',m')}$ between two lines $L_{(e,m)}$ and $L_{(e',m')}$ is determined by their self spins $\theta_{(e,m)}$ as $B_{(e,m),(e',m')}=\theta_{(e+e',m+m')}/(\theta_{(e,m)}\theta_{(e',m')})$. However, in order to contrast with the analogue condition in higher dimensions, we still present the second property explicitly. The third property is guaranteed by the fact that there are $N$ lines in $\CA$, which we will verify at the end of this subsection.

To enumerate all possible Lagrangian subgroups, we first assume that a particular line $L_{(e,m)}$ belongs to $\CA$, hence $L_{(ke, km)}$ also belongs to $\CA$ due to group structure, for any $k\in \Z$.

The trivial topological spin means $e$ and $m$ can not be simultaneously coprime with $N$. This means that there must exist a $k<N$ such that $L_{(ke, km)}$ is a purely electric line $L_{(p',0)}$ or purely magnetic line $L_{(0,q')}$. 
So any Lagrangian subgroup contains at least one non-trivial purely electric or magnetic line. 
Without loss of generality, we assume that a electric line $L_{(p',0)}$ belongs to $\CA$, where $1<p'<N$.  Under multiplication,  $L_{(p,0)}$ also belongs to $\CA$, where $p= \gcd(p',N)$. Note that $L_{(p,0)}$ is the purely electric line with the \emph{smallest} electric charge. For convenience, we also denote $q=N/p$. Summarizing the above, we have shown that $\CA$ contains $q$ purely electric lines among $N$ lines in total
\begin{eqnarray}\label{eq:elines}
    L_{(kp,0)}, \qquad k=0, 1, ..., q-1. 
\end{eqnarray}

Suppose $L_{(e,m)}$ also belongs to $\CA$ (which is independent of the $L_{(e,m)}$ in the previous paragraph), with $m\neq 0$ mod $N$. Trivial mutual braiding with purely electric lines requires $pm=0$ mod $N$, or equivalently
\begin{eqnarray}
    m= s q
\end{eqnarray}
for \emph{certain} $s\in \Z$. Denoting $t=\gcd(s,p)$, it follows that under multiplication, $L_{(e', tq)}$ also belongs to $\CA$, for certain $e'$. 
Trivial topological spin of $L_{(e',tq)}$ requires 
\begin{eqnarray}\label{eq:te'}
    te'= 0 \text{ mod } p.
\end{eqnarray}
Note that $tq$ is the minimal magnetic charge mod $N$ in the orbit $L_{(ke, ksq)}$ for different $k\in \Z$. Supplementing $L_{(e',tq)}$ to the set \eqref{eq:elines}, we find the following lines in the Lagrangian subgroup $\CA$, 
\begin{eqnarray}
    L_{(kp+k'e', k'tq)}, \qquad k=0, 1, ..., q-1, \qquad k'=0,1,...,p/t-1.
\end{eqnarray}
Thus we find $qp/t=N/t$ lines in the Lagrangian subgroup. Since the Lagrangian subgroup contains $N$ lines, unless $t=1$, the above set does not contain enough lines to form a Lagrangian subgroup. However, since the charge lattice is two dimensional, any missing line should be generated by the two generators $L_{(p,0)}$ and $L_{(e', tq)}$ (note that they are not linearly dependently when $t>1$), which is a contradiction. This implies that $t=1$, and by \eqref{eq:te'}, $e'=0$ mod $p$.\footnote{Another way to see $t=1$ is as follows. We prove by contradiction. Assume $t>1$, and we would like to find an operator that is not generated by $L_{(p,0)}$ and $L_{(e', tq)}$. By \eqref{eq:te'}, $e'= (p/t)x$ for $x\in \Z$. When $x=0$ mod $t$, $e'=0$ mod $p$, and one can compose $L_{(e',tq)}$ with $L_{(p,0)}$ to get $L_{(0,tq)}$. Clearly, the generator $L_{(0,q)}$ can not be generated by $L_{(p,0)}$ and $L_{(0,tq)}$. When $x\neq 0$ mod $t$, we can assume $0<x<t$, i.e. $0<e'<p$ without loss of generality. We can then consider $L_{(0,tq)}$, which has the trivial topological spin and the trivial mutual braiding with both $L_{(p,0)}$ and $L_{(e',tq)}$ thanks to \eqref{eq:te'}. Because $0<e'<p$, $L_{(0,tq)}$ is not generated by $L_{(p,0)}$ and $L_{(e',tq)}$. In both cases, we find at least one operator that can be added into the Lagrangian subgroup, showing that $L_{(p,0)}$ and $L_{(e',tq)}$ do not generate the full Lagrangian subgroup.  When $t=1$ however, \eqref{eq:te'} shows $e'=0$ mod $p$, and clearly $L_{(p,0)}$ and $L_{(e',q)}$ generate the entire Lagrangian subgroup.} In conclusion, the Lagrangian subgroup is completely specified by $p$ where $p|N$, which is generated by $L_{(p,0)}$ and $L_{(0,N/p)}$, i.e. 
\begin{eqnarray}
    \CA_p= \{L_{(xp, yN/p)}|x\in \Z_{N/p}, y\in \Z_p\}.
\end{eqnarray}

Finally, we show the third property of the Lagrangian subgroup, that for any line $L_{(e,m)}$ not within $\CA_p$, it must braid non-trivially with at least one element in $\CA_p$. We prove by contradiction. Suppose there exists $L_{(e,m)}\notin \CA_p$ which braids trivially with every element in $\CA_p$. The assumption implies $eyN/p + mxp=0$ mod $N$ for any $x,y$. We first take $y=0$, then $mxp=0$ mod $N$ for any $x$ implies $m= 0$ mod $N/p$. Similarly, by taking $x=0$, then $eyN/p=0$ mod $N$ for any $y$ implies $e=0$ mod $p$. Thus $L_{(e,m)}\in \CA_p$, which contradicts with the assumption. This completes the proof.

\subsection{$\Z_2^{\text{em}}$ stable Lagrangian subgroup}
\label{sec:3dstable}

We further classify which topological boundary condition is $\Z_2^{\text{\text{em}}}$ stable. The $\Z_2^{\text{\text{em}}}$ symmetry \eqref{eq:emsymmetry} maps the charges $(e,m)$ to $(vm, ue)$. The generators of the Lagrangian subgroup $\CA_p$ are mapped to 
\begin{eqnarray}
    L_{(p,0)}\to L_{(0,up)}, \qquad L_{(0,N/p)} \to L_{(vN/p,0)}.
\end{eqnarray}
$\Z_2^{\text{\text{em}}}$ stability implies the above $\Z_2^{\text{\text{em}}}$ of the generators also belong to $\CA_p$, i.e. 
\begin{eqnarray}
    L_{(0,up)} = L_{(xp, yN/p)}, \qquad L_{(vN/p,0)} = L_{(zp, wN/p)}.
\end{eqnarray}
The above implies $x\in (N/p)\Z$, $w\in p\Z$, and 
\begin{eqnarray}\label{eq:pnp}
    up= yN/p \text{ mod } N, \qquad  vN/p=zp \text{ mod }N.
\end{eqnarray}
Combining the two conditions, we find $zy=1$ mod $p$ as well as $zy=1$ mod $N/p$. This in particular implies $y$ and $z$ are coprime with both $p$ and $N/p$. On the other hand, the first condition in \eqref{eq:pnp} implies $p=(vy+p\alpha)N/p$ for certain integer $\alpha$. However, because $y$ is coprime with $p$ and $v$ is coprime with $N$ (hence $p$), $(vy+p\alpha)$ is also coprime with $p$. So this condition can be satisfied only when $(N/p)|p$. Similarly, the second equality in \eqref{eq:pnp} implies $p|(N/p)$. Combining the two conditions, we find 
\begin{eqnarray}
    p=N/p
\end{eqnarray}
which means $N=p^2$ is a perfect square. By the results in Sec.~\ref{sec:criterias}, we conclude that the $\Z_N^{(0)}$ duality defect is group theoretical if and only if $N$ is a perfect square, for any choice of $u,v$ (i.e. the bicharacter). Since the choice of the Frobenius-Schur indicator $\epsilon$ does not enter the above discussion, the group-theoretical condition is also independent of  $\epsilon$ as well. We emphasize that this result has been already proven in \cite{gelaki2009centers}, and hope that our discussion is more accessible to physicists.

\subsection{SymTFT as a Dijkgraaf-Witten theory}
\label{sec:3dDW}
In Sec.~\ref{sec:3dstable}, we showed that a $\Z_N$ duality defect is group theoretical if and only if $N$ is a perfect square, for any choice of $u,v,\epsilon$. In this section, we would like to explicitly show that the SymTFT for a perfect square $N$ is indeed a Dijkgraaf-Witten theory.

Since $N=p^2$, the $\Z_N$ cochains $a,\widehat{a}$ can be rewritten as 
\begin{eqnarray}\label{eq:decomposition}
    a= p \widehat{b} +c', \qquad \widehat{a}= p \widehat{c}+b'
\end{eqnarray}
where $\widehat{c}$, $\widehat{b}$ are $\Z_p$ cochains, and $b'$, $c'$ are $\Z_N$ cochains. In terms of these new variables, \eqref{eq:ZNaction} is rewritten as 
\begin{eqnarray}\label{eq:Lprime}
	\CL=\frac{2\pi}{N} (p\widehat{c}+ b')\delta (p \widehat{b}+ c') = \frac{2\pi}{p} b\delta \widehat{b} + \frac{2\pi}{p} \widehat{c}\delta c + \frac{2\pi}{N} b'\delta c'.
\end{eqnarray} 
In the second equality, we dropped $2\pi \widehat{b}\delta \widehat{c}$ since it belongs to $2\pi \Z$.  On the right hand side, the first two terms are the standard BF couplings where $b=b'$ mod $p$ and $c=c'$ mod $p$, while the last term is the DW twist term. Thus the $\Z_N$ gauge theory without a DW term is equivalent to $\Z_p\times \Z_p$  gauge theory with a DW term.

The advantage of recasting \eqref{eq:ZNaction} into \eqref{eq:Lprime} is that the electric-magnetic exchange symmetry exchanging $a$ and $\widehat{a}$ becomes a symmetry that only exchanges among the electric fields $b,c$, and among the magnetic fields $\widehat{b}, \widehat{c}$, separately. But the electric and magnetic fields do not mix under $\Z_2^{\text{\text{em}}}$. 
Concretely, the \eqref{eq:emsymmetry} acts on the $\Z_p$ cochains via 
\begin{eqnarray}\label{eq:Zem2}
    b \to  v c, \qquad c \to  u b, \qquad \widehat{b} \to  u \widehat{c}, \qquad \widehat{c} \to  v \widehat{b}. 
\end{eqnarray}
Thus gauging \eqref{eq:emsymmetry} of \eqref{eq:ZNaction} is equivalent to gauging \eqref{eq:Zem2} of \eqref{eq:Lprime}. From the latter, it is almost by definition that the gauged theory is Dijkgraaf-Witten, where the definition is reviewed in Sec.~\ref{sec:intro2}.

We remark that the new variables introduced in \eqref{eq:decomposition} is motivated from the Lagrangian subgroup derived in Sec.~\ref{sec:3dstable}. The Lagrangian subgroup is generated by $e^{\frac{2\pi i}{N} p \oint a}= e^{\frac{2\pi i}{p} \oint c}$ and $e^{\frac{2\pi i}{N} p \oint \widehat{a}}= e^{\frac{2\pi i}{p} \oint b}$, and we require that these operators are closed under $\Z_2^{\text{em}}$ transformation. This is obvious from \eqref{eq:Zem2}.

Let's derive the Lagrangian from gauging \eqref{eq:Zem2} of \eqref{eq:Lprime}. The first step is to couple to the $\Z_2^{\text{em}}$ background field $x$, and sum over all the flat configuration of $x$, i.e. gauge $\Z_2^{\text{em}}$. Coupling to the $\Z_2^{\text{em}}$ background field amounts to changing the ordinary differential operator $\delta$ to the twisted differential operator $\delta_x$, and ordinary cup product $\cup$ to twisted cup product $\cup_x$. See \cite[App.~A]{Benini:2018reh} for a review of twisted cochains, differentials and cup products. In components, the $\Z_2^{\text{em}}$ gauged Lagrangian on a 3-simplex $(ijkl)$ is 
\begin{equation}\label{eq:twistcocycle}
\begin{split}
    \CL^{\text{gauged}}_{ijkl} = \frac{2\pi}{p} \widehat{\mathbf{b}}_{ij}^{T} K^{x_{ij}} \left(K^{x_{jk}} \mathbf{b}_{kl} - \mathbf{b}_{jl} + \mathbf{b}_{jk}\right) +\frac{\pi}{N} {\mathbf{b}'}_{ij}^T \sigma^x K^{x_{ij}} \left(K^{x_{jk}} \mathbf{b}'_{kl} - \mathbf{b}'_{jl} + \mathbf{b}'_{jk}\right)+ \pi \epsilon x_{ij} x_{jk}x_{kl}
\end{split}
\end{equation}
where 
\begin{eqnarray}
    K= 
    \begin{pmatrix}
        0 & v\\
        u  & 0
    \end{pmatrix}, \qquad \mathbf{b}= 
    \begin{pmatrix}
        b\\
        c
    \end{pmatrix}, \qquad 
    \mathbf{b}'= 
    \begin{pmatrix}
        b'\\
        c'
    \end{pmatrix}
\end{eqnarray}
and $x$ is the dynamical, flat, $\Z_2^{\text{em}}$ gauge field. The last term is the twist one can add upon gauging $\Z_2$, whose coefficient $\epsilon$ is related to the FS indicator $(-1)^\epsilon$. The action is invariant under the gauge transformations 
\begin{equation}
	\begin{split}
		\mathbf{b}_{ij} \to K^{-\gamma_i} (\mathbf{b}_{ij} +K^{x_{ij}} \mathbf{\beta}_{j} - \mathbf{\beta}_i), \quad \widehat{\mathbf{b}}_{ij} \to (K^T)^{-\gamma_i} (\widehat{\mathbf{b}}_{ij} +K^{x_{ij}} \widehat{\mathbf{\beta}}_{j} - \widehat{\mathbf{\beta}}_i), \quad x_{ij} \to x_{ij} + \gamma_j - \gamma_i
	\end{split}.
\end{equation}
Summing over $\widehat{\mathbf{b}}$ constrains ${\mathbf{b}}$ to be a twisted cocycle (i.e. it is twisted-flat), in components, $(\delta_x \mathbf{b})_{jkl}= K^{x_{jk}} \mathbf{b}_{kl} - \mathbf{b}_{jl} + \mathbf{b}_{jk}=0$ mod $p$. Thus the partition function is independent of how the $\Z_p$ fields $\mathbf{b}$ are lifted to $\Z_N$ fields $\mathbf{b}'$. The full partition function is
\begin{eqnarray}
	\CZ_{\text{gauged}}= \sum_{(\mathbf{b}, x)} \prod_{ijkl} \exp\left(\frac{\pi i}{N} \mathbf{b}_{ij}^T \sigma^x K^{x_{ij}} \left(K^{x_{jk}} \mathbf{b}_{kl} - \mathbf{b}_{jl} + \mathbf{b}_{jk}\right)+ \pi i \epsilon x_{ij} x_{jk}x_{kl}\right).
\end{eqnarray} 
The pair $(\mathbf{b}, x)$ with the above flatness condition is the gauge field for the gauge group $(\Z_p\times \Z_p)\rtimes \Z_2$, with the $\Z_2$ exchanging the two $\Z_p$'s.\footnote{The special case of $N=4, p=2$ was discussed in a lattice model in \cite{2023PhRvL.130b6801Z}.} The partition function is clearly of the form of DW, for any choice of $u,v, \epsilon$.

\subsection{Explicit topological manipulation}
\label{sec:3dtopomanipulation}

We finally proceed to find an explicit topological manipulation that maps the duality defect to an invertible defect.

Suppose a 2d QFT $\CX$ is self-dual. This means
\begin{eqnarray}\label{eq:SNX}
	\CZ_{\CX}[A] = \sum_{a\in \Z_N } \CZ_{\CX}[a] e^{\frac{2\pi i v }{N}\int_{M_2} aA}.
\end{eqnarray}
The summation $a\in \Z_N$ means summing over 1-cochain $a$ valued in $\Z_N$, with flatness condition $\delta a= 0$ mod $N$,  modulo gauge transformations. The parameter $v$ (which is coprime with $N$) paramaterizes different ways of gauging $\Z_N$.  By gauging $\Z_N$ on half of the spacetime, we get a duality defect $\CN$.

To motivate the topological manipulation, we first note that the QFT $\CX$ corresponds the SymTFT being $\Z_N$ gauge theory and Dirichlet boundary condition of $a$. Under the decomposition \eqref{eq:decomposition}, the Dirichlet boundary condition of $a$ translates to the Dirichlet boundary condition of $c$ and Dirichlet boundary condition of $\widehat{b}$. Given a Dijkgraaf-Witten theory as the SymTFT, the invertible symmetry corresponds to the Dirichlet boundary condition of electric fields i.e. $b,c$ only, so the topological manipulation should transform the above topological boundary condition to the Dirichlet boundary condition of both $b$ and $c$. Concretely, in terms of the boundary states defined in Sec.~\ref{sec:symtftintro}, 
\begin{eqnarray}\label{eq:newtopobdystate}
    \bra{A} \to \sum_{\widehat{b}\in \Z_p}  \bra{p\widehat{b} + C} e^{\frac{2\pi i}{N} \int_{X_2} \widehat{b} B}
\end{eqnarray}
where $\bra{A}$ is the Dirichlet boundary state with $\Z_N= \Z_{p^2}$ background field $A$, the second factor on the RHS is the boundary term coming from the integration by parts of the first term in \eqref{eq:Lprime}, so that it has the standard BF coupling $\frac{2\pi}{p} \widehat{b} \delta b$. We also set the electric fields $b,c$ in the SymTFT \eqref{eq:Lprime} to background fields $B, C$ respectively. Taking the inner product between the new topological boundary state \eqref{eq:newtopobdystate} and the dynamical boundary state \eqref{eq:dynbdystate}, we get the new partition function 
\begin{eqnarray}\label{eq:SpX}
    \CZ_{\widetilde{\CX}}[B,C]= \sum_{\widehat{b}\in \Z_p, \delta \widehat{b}= - \beta C}\CZ_{\CX}[p\widehat{b}+C] e^{\frac{2\pi i}{p}\int_{M_2} \widehat{b}B}.
\end{eqnarray}
The summation is over all $\widehat{b}$ with the constraint $\delta \widehat{b} = - \beta C$, modulo gauge transformations.\footnote{$\beta$ is the Bockstein homomorphism $\beta: H^1(M_2, \Z_p)\to H^2(M_2, \Z_p)$.} The constraint comes from the flatness condition of $A$ before topological manipulation, which descends to the flatness condition of $p\widehat{b}+C$. The constraint also enforces that the exponent $\frac{2\pi i }{p}\int_{M_2}\widehat{b}B$ is not gauge invariant, and one way to make it gauge invariant is to introduce a 3d bulk\cite{Tachikawa:2017gyf}, 
\begin{eqnarray}\label{eq:3dspt}
	\frac{2\pi i }{p}\int_{M_3} B \beta C.
\end{eqnarray}
This shows that after gauging, there is a mixed anomaly between two $\Z_p$ symmetries.

We claim that \eqref{eq:SpX} defines the desired topological manipulation mapping the self duality under gauging $\Z_N$ to an invertible symmetry. To see this, we perform a self-duality transformation on the right hand side of \eqref{eq:SpX}. Concretely, we substitute \eqref{eq:SNX} to the right hand side of \eqref{eq:SpX}, and get
\begin{eqnarray}\label{eq:SpXBC}
	\begin{split}
		\CZ_{\widetilde{\CX}}[B,C] &= \sum_{\substack{a\in \Z_N, \widehat{b}\in \Z_p, \\\delta \widehat{b}= - \beta C}} \CZ_{\CX}[a] e^{\frac{2\pi i }{N}\int_{M_2} p\widehat{b}B +v a(p\widehat{b}+C)}.
	\end{split}
\end{eqnarray}
Summing over $\widehat{b}$ enforces $a=u B$ mod $p$, which is equivalent to $a=pc+uB$ mod $N$. The flatness condition of $a$ enforces $\delta c = -u\beta B$. Substituting this solution to the partition function, the exponent becomes $\frac{2\pi i}{N}\int_{M_2} v(pc+u B)C= \frac{2\pi i}{p}\int_{M_2}v cC + \frac{2\pi i}{N}BC$. The last term is a counter term, which can be absorbed to the bulk anomaly SPT action \eqref{eq:3dspt} by exchanging $B\leftrightarrow C$, i.e. $ \frac{2\pi i }{N}\int_{M_2}BC +  \frac{2\pi i}{p}\int_{M_3}B\beta C =  \frac{2\pi i}{p}\int_{M_3}C\beta B$. Combining the above, \eqref{eq:SpXBC} finally becomes 
\begin{eqnarray}
	\CZ_{\widetilde{\CX}}[B,C] = \sum_{c\in \Z_p, \delta c = -u \beta B} \CZ_{\CX}[pc+uB] e^{\frac{2\pi i}{p}\int_{M_2} v cC} = \CZ_{\widetilde{\CX}}[vC,u B].
\end{eqnarray}
This shows that the self duality symmetry in $\CX$ is mapped to an invertible $\Z_2$ symmetry that simply maps the background fields as 
\begin{eqnarray}
    B\to v C, \qquad C\to uB
\end{eqnarray}
which is indeed consistent with the transformation \eqref{eq:Zem2} in the SymTFT.

\section{Group theoretical duality defects in 4d}
\label{sec:3+1}

We generalize the discussion of duality defects in 2d to duality defects in 4d. The discussion is overall in parallel with Sec.~\ref{sec:1+1}, but there are additional subtleties which we highlight. 

\subsection{Lagrangian subgroups of 5d $\Z_N$ 2-form gauge theory}

The action of the $\Z_N$ 2-form gauge theory is 
\begin{eqnarray}\label{eq:5dBF}
    \CL= \frac{2\pi}{N} \widehat{b}\delta b,
\end{eqnarray}
where $b, \widehat{b}$ are both $\Z_N$ 2-cochains. It has a $\Z_4^{\text{em}}$ exchange symmetry 
\begin{eqnarray}\label{eq:Z4em}
    b\to u\widehat{b}, \qquad \widehat{b}\to -vb,
\end{eqnarray}
where $uv=1$ mod $N$. 
Comparing with \eqref{eq:emsymmetry}, the additional minus sign comes from the change of form degree of the gauge fields, which consequently makes the exchange symmetry to be order four, rather than order two.\footnote{When $N=2$, due to $b=-b$ mod $2$, $\Z_4^{\text{em}}$ reduces to $\Z_2^{\text{em}}$. } The topological surfaces are 
\begin{eqnarray}
    S_{(e,m)}(\sigma)= e^{\frac{2\pi i e}{N}\oint_{\sigma} b} e^{\frac{2\pi i m}{N}\oint_{\sigma} \widehat{b}}.
\end{eqnarray}
One significance is that the topological surface operator can be modified by a local counter term $e^{\frac{2\pi i k}{2N}\CP(\sigma)}$ for even $N$, and $e^{\frac{2\pi i}{N}\frac{N+1}{2}\braket{\sigma,\sigma}}$ for odd $N$, where we $\CP(\sigma)$ is the Pontryagin square of $\sigma\in H_2(X,\Z_N)$. In the absence of such counter term, the surface operators are not closed under fusion, $S_{(e,m)}(\sigma) S_{(e',m')}(\sigma) = e^{\frac{2\pi i}{N} me'\braket{\sigma,\sigma}} S_{(e+e',m+m')}(\sigma)$.

A full classification of topological boundary conditions of a generic 5d TQFT is still under development. However, for 5d $\Z_N$ 2-form gauge theory, its topological boundary conditions are expected to be classified by the Lagrangian subgroup of its surface operators, with additional data specifying the topological refinements \cite{Chen:2021xuc}. We will take this as an assumption throughout the rest of the paper.  In particular, the surface operators within the Lagrangian subgroup are closed under fusion.\footnote{One way to understand the closedness under fusion is as follows. Assuming two operators $\widetilde{S}_{(e,m)}$ and $\widetilde{S}_{(e',m')}$ are within the Lagrangian subalgebra, where $\widetilde{S}_{(e,m)}$ is related to $S_{(e,m)}$ by stacking a counter term specified in the previous paragraph.   This means that the associated boundary state $\ket{\CB}$ is stabilized by both of them. $\widetilde{S}_{(e,m)}\ket{\CB}=\ket{\CB}$, $\widetilde{S}_{(e',m')}\ket{\CB}=\ket{\CB}$. This means that their product $\widetilde{S}_{(e,m)}\widetilde{S}_{(e',m')}$ also stabilizes the boundary state $\ket{\CB}$, hence belongs to the Lagrangian subalgebra. Now, let $(e+e',m+m')=(0,0)$ mod $N$, hence $\widetilde{S}_{(e,m)}\widetilde{S}_{(e',m')}$ is at most a phase specified by the counter term. If the counter term is non-trivial, the boundary state must vanish. Thus we should carefully choose the counter term such that  $\widetilde{S}_{(e,m)}\widetilde{S}_{(e',m')}= 1$ whenever $(e+e',m+m')=(0,0)$ mod $N$. This is ensured if we demand the operators $\widetilde{S}_{(e,m)}$ are closed under fusion for arbitrary $(e,m)$, i.e. $\widetilde{S}_{(e,m)}\widetilde{S}_{(e',m')}=\widetilde{S}_{(e+e',m+m')}$. }
The Lagrangian subgroup $\CA$ consists of $N$ topological surface operators $S_{(e,m)}$ with the following conditions, 
\begin{enumerate}
    \item $S_{(e,m)}$ has trivial topological spin. This is automatically satisfied for any $(e,m)$ \cite{Chen:2021xuc}.
    \item Any two surfaces $S_{(e,m)}$ and $S_{(e',m')}$ in the Lagrangian subgroup $\CA$ have trivial mutual braiding, i.e. $e^{\frac{2\pi i}{N}(em'-me')}=1$. 
    \item Any other surface operator that does not belong to $\CA$ braids non-trivially with at least one line in $\CA$. 
\end{enumerate}
Since the trivial self-braiding condition is automatically satisfied, the Lagrangian subgroup in 5d is less constrained than that in 3d, hence the structure of $\CA$ is richer. For example, we will see that the Lagrangian subgroups can be generated by an arbitrary single surface $\widetilde{S}_{(e,m)}$ as long as $\gcd(e,m,N)=1$, while in the 3d $\Z_N$ gauge theory we should further require it to be a self-boson $em=0$ mod $N$ which is much more constraining.

Like in 3d $\Z_N$ gauge theory, the third condition is guaranteed when there are $N$ topological surface operators in $\CA$. We will verify this at the end of this subsection.

Below, we classify the topological boundary condition of $\Z_N$ 2-form gauge theory by classifying its Lagrangian subgroups. We first derive the classification at the level of charges in Sec.~\ref{sec:Lag1}, by assuming that the surface operators within it are closed fusion, i.e. 
\begin{eqnarray}
        \widetilde{S}_{(e,m)}(\sigma) \widetilde{S}_{(e,m)}(\sigma') &=& \widetilde{S}_{(e,m)}(\sigma+\sigma')\label{eq:fusion1}, \\
        \widetilde{S}_{(e,m)}(\sigma) \widetilde{S}_{(e',m')}(\sigma) &=& \widetilde{S}_{(e+e',m+m')}(\sigma) \label{eq:fusion2},
\end{eqnarray}
where $\widetilde{S}_{(e,m)} = S_{(e,m)}e^{i K_{e,m}\braket{\sigma,\sigma}}$ with $K_{e,m}$ determined by the closeness under fusion. We then discuss the phase factors in Sec.~\ref{sec:Lag2}, where we will find for each charge $(e,m)$ there can be distinct topological refinements, as emphasized in \cite{Chen:2021xuc}.

\subsubsection{Classifying Lagrangian subgroup: charges}
\label{sec:Lag1}

We first classify the charges of the surface operators in the Lagrangian subgroup, following the discussion in Sec.~\ref{sec:tbc3d}. In App.~\ref{app:alternative} we provide an alternative derivation of the Lagrangian subgroups.

Suppose there are pure electric surface operators in the Lagrangian subgroup, and the one with \emph{minimal} electric charge is $\widetilde{S}_{(p,0)}$, where $p|N$ and $1\leq p\leq N$.\footnote{ When $p=N$, there is no pure electric operator.}  It is clear that there are additional operators  $\widetilde{S}_{(\ell',s N/p)}$ having trivial mutual braiding with $\widetilde{S}_{(p,0)}$, hence can be supplemented into the Lagrangian subgroup. Denoting $t=\gcd(s,p)$, it follows that $\widetilde{S}_{(\ell,t N/p)}$ also belongs to the Lagrangian subgroup, where $\ell=x\ell'$ mod $p$ and $sx=t$ mod $p$.

We proceed to show that $t=1$ for $\widetilde{S}_{(p,0)}$ and $\widetilde{S}_{(\ell, t N/p)}$ to generate the entire Lagrangian subgroup. To see this, we denote $\gcd(t,\ell)=\tilde{\ell}$.  It follows that $\widetilde{S}_{(\ell/\tilde{\ell}, t N/p\tilde{\ell})}$ also has trivial braiding with the above two generators, and should belong to the Lagrangian subgroup, meaning that it can be expended by the generators.  This is possible only when $\tilde{\ell}=1$, meaning $t$ and $\ell$ are coprime. Then we fuse $p/t$ copies of the second generator to get $\widetilde{S}_{(\ell p/t, 0)}$, hence should be generated by the first generator, i.e. $\ell p/t = p x$ for an integer $x$.  This means $\ell= xt$, in other words, $\gcd(\ell, t)= t=1$, as desired.  When $t=1$, indeed $\widetilde{S}_{(p,0)}$ and $\widetilde{S}_{(\ell,  N/p)}$ generate $(N/p)\times p= N$ surfaces, which is the size of the Lagrangian subgroup.
In summary, Lagrangian subgroups are specified by $(\ell, p)$, generated by $\widetilde{S}_{(p,0)}$ and $\widetilde{S}_{(\ell,  N/p)}$, i.e. 
\begin{eqnarray}
    \CA_{(\ell, p)}= \{\widetilde{S}_{(xp+y\ell, yN/p)}|x\in \Z_{N/p}, y\in \Z_{p}\}.
\end{eqnarray}

We finally verify the third property of the Lagrangian subgroup, that for any surface $S_{(e,m)}$ not within $\CA_{(\ell,p)}$, it must braid non-trivially with at least one element in $\CA_{(\ell,p)}$. We again prove by contradiction. Suppose there exists $S_{(e,m)}\notin \CA_{(\ell, p)}$ which braids trivially with every element in $\CA_{(\ell,p)}$. The assumption means
\begin{eqnarray}
    ey\frac{N}{p} - m(xp+y\ell)=0 \text{ mod }N, \qquad \forall x\in \Z_{N/p},y\in \Z_{p}.
\end{eqnarray}
Setting $y=0$, we get $mxp=0$ mod $N$ for any $x\in \Z_{N/p}$, which means $m=0$ mod $N/p$. Let's denote $m= \widehat{m} N/p$. We can alternatively set $x=0$, then $(eN/p - m \ell)y=0$ mod $N$ for any $y$, meaning $e-\widehat{m} \ell=0$ mod $p$. Let's then denote $e= p \widehat{e} + \ell \widehat{m}$. So 
\begin{eqnarray}
    (e,m)= (p \widehat{e} + \ell \widehat{m}, \widehat{m} N/p) = (p,0) \widehat{e} + (\ell, N/p) \widehat{m} \in \CA_{(\ell,p)}.
\end{eqnarray}
This contradicts with the assumption that $(e,m)$ is not within $\CA_{(\ell,p)}$. Hence the third property in the Lagrangian subgroup is verified.

\subsubsection{Classifying Lagrangian subgroup: including phases}
\label{sec:Lag2}

In Sec.~\ref{sec:Lag1}, we have determined the charges in the Lagrangian subgroups. In this section, we explicitly construct the surface operators, in particular specify the counter terms within $\widetilde{S}_{(e,m)}$. The most general ansatz is 
\begin{eqnarray}\label{eq:ansatz}
    \widetilde{S}_{(e,m)}(\sigma) = S_{(e,0)}(\sigma) S_{(0,m)}(\sigma) e^{i K_{e,m}\braket{\sigma,\sigma}}.
\end{eqnarray}
Closedness under fusion requires \eqref{eq:fusion1} and \eqref{eq:fusion2}. 
In the above, $(e,m)$ belongs to $k(p,q)$ for one generator case, and $(px+\ell y, yN/p)$ for two generators case. We discuss these two cases separately.

We substitute $(e,m)= (xp+y\ell, yN/p)$ with $x=0,1,...,N/p-1$ and $y=0,1,...,p-1$. Similarly, $(e',m')= (x'p+y'\ell, y'N/p)$. However, $(e+e',m+m')=((x+x')p+(y+y')\ell, (y+y')N/p)$ should be more precisely written as 
\begin{eqnarray}
    (e',m')=[x+x'+ (y+y'-[y+y']_p)\frac{\ell}{p}]_{N/p}(p,0)+ [y+y']_p(\ell,  N/p),
\end{eqnarray}
where by $[x]_p$ is the mod $p$ function taking value in $0,\cdots,p-1$.
Then \eqref{eq:fusion1} yields
\begin{eqnarray}\label{eq:K2}
    K_{xp+yLr, yN/p}= 
    \begin{cases}
        \frac{2\pi}{N}\frac{N+1}{2}(xp+y\ell)yN/p \text{ mod } 2\pi, & N \text{ odd }\\
        \frac{2\pi}{2N}(xp+y\ell)yN/p + \pi J_{x,y} \text{ mod } 2\pi , & N \text{ even } \\
    \end{cases}.
\end{eqnarray}
Substituting \eqref{eq:K2} into \eqref{eq:fusion2}, we find that for odd $N$ \eqref{eq:K2} is automatically satisfied, while for even $N$, $J_{x,y}$ should satisfy 
\begin{equation}\label{eq:longeq}
\begin{split}
    &\frac{\ell}{p}((y+y')^2-[y+y']_p^2)+ xy+x'y'-[x+x'+(y+y'-[y+y']_p)\ell/p]_{N/p} [y+y']_p\\
    &=J_{[x+x'+(y+y'-[y+y']_p)\ell/p]_{N/p}, [y+y']_p}- J_{x,y}-J_{x',y'} \text{ mod } 2.\\
\end{split}
\end{equation}

It is not straightforward to explicitly solve for $J_{x,y}$ analytically. However, by numerically solving \eqref{eq:longeq} for $N\leq 100$, we confirmed that the solution exist for all even $N$'s: there are four solutions when $p, N/p, \ell$ are all even, and two solutions otherwise. In particular, from the numerical results, we find that when $p, N/p, \ell$ are all even, the four solutions are of the form $J_{x,y}= xy + c_1 x + c_2 y$ mod $2$, where $c_1, c_2=0,1$ parameterize the four solutions. Another situation where we are able to solve $J_{x,y}$ is when the Lagrangian subgroup is generated by only one surface operator, which we will discuss in App.~\ref{app:onegen}.

The situation significantly simplifies when we restrict the 4d boundary of the 5d SymTFT to be spin manifolds. In this case,   the $\pi J_{x,y}$ term in \eqref{eq:K2} can be ignored due to $\braket{\sigma,\sigma}= 0$ mod $2$.

\subsection{$\Z_4^{\text{em}}$ stable Lagrangian subgroup}
\label{sec:5dstableLag}

We proceed to examine when the Lagrangian subalgebra is stable under $\Z_4^{\text{em}}$. Recall that under $\Z_4^{\text{em}}$, the gauge field transforms as \eqref{eq:Z4em}.  So $S_{(e,0)}(\sigma)\to S_{(0,ue)}(\sigma)$, and $S_{(0,m)}(\sigma) \to S_{(-v m, 0)}(\sigma)$.

When the 4d spacetime $X_4$ is a spin manifold, the counter terms for even and odd $N$ are given by \eqref{eq:K2} with $J_{x,y}=0$. Under $\Z_4^{\text{em}}$, the generators $\widetilde{S}_{(p,0)}$ and $\widetilde{S}_{(\ell, N/p)}$ are mapped to $\widetilde{S}_{(0,up)}$ and $\widetilde{S}_{(-vN/p, u\ell)}$ respectively for both even and odd $N$. Stability under $\Z_4^{\text{em}}$ implies $(0,up) = (xp+y \ell, yN/p)$ mod $N$, and $(-vN/p,u\ell) = (zp+w\ell, wN/p)$ mod $N$. In components, we have 
\begin{eqnarray}
    xp+ y\ell &=& 0 \text{ mod } N, \label{eq:oddNtwogen1}\\
    yN/p &=& up \text{ mod } N, \label{eq:oddNtwogen2}\\
    zp+w\ell &=& -vN/p \text{ mod } N, \label{eq:oddNtwogen3}\\
    wN/p &=& u\ell \text{ mod } N.\label{eq:oddNtwogen4}
\end{eqnarray}
Once the $\Z_4^{\text{em}}$ images of the generators are within the Lagrangian subgroup, it is not hard to show that the $\Z_4^{\text{em}}$ images of all other surface operators in the Lagrangian subgroup are all within the Lagrangian subgroup as well.

Given $N,u,v$, can we find $p, \ell, x,y,z,w$ such that \eqref{eq:oddNtwogen1}$\sim$\eqref{eq:oddNtwogen4} are satisfied? Below, we provide an equivalent but simpler-looking criteria of $N$ for which the solutions exists.\footnote{We are grateful to Justin Kaidi for discussing the following proof.} First, multiplying $p$ on both sides of \eqref{eq:oddNtwogen2} yields $p^2=0$ mod $N$ (note that $u,v$ are coprime with $N$). This means
\begin{eqnarray}\label{eq:p2NM}
    p^2=NM
\end{eqnarray}
for some integer $M$.  As a consequence, we have $p= (N/p)M$, which means
\begin{eqnarray}
    L:=\gcd(p, N/p)= N/p,
\end{eqnarray}
hence we write $p=ML$. Substituting this into \eqref{eq:p2NM}, we have $(ML)^2= NM$, which gives rise to
\begin{eqnarray}
    N= L^2 M.
\end{eqnarray}
So far, we only used \eqref{eq:oddNtwogen2}. 
To see the condition of $M$, we multiply $\ell$ on both sides of \eqref{eq:oddNtwogen4}, 
\begin{eqnarray}
    \ell^2= v w\ell N/p = v(-vN/p- zp)N/p = -(vN/p)^2 = -v^2 L^2 \text{ mod } N
\end{eqnarray}
where in the second equality we used \eqref{eq:oddNtwogen3}. Multiplying $u^2$ and dividing $L^2$ on both sides, we find
\begin{eqnarray}
    (u\ell/L)^2= -1 \text{ mod } M.
\end{eqnarray}
In other words, $\ell=Lrv$ with $r^2=-1$ mod $M$. Finally, we note that \eqref{eq:oddNtwogen1} does not impose further constraints and can be solved by $x=-r$ and $y=uM$.

So far,  we have shown that \eqref{eq:oddNtwogen1}$\sim$\eqref{eq:oddNtwogen4} imply $N=L^2M$ where $-1$ is a quadratic residue of $M$. To show the converse is also true, we simply take  
\begin{eqnarray}
    (N, x,y,z,w,p, \ell)=(L^2M, -r, uM, -v\lambda, r, LM, Lrv)
\end{eqnarray}
where $r$ satisfies $r^2+1= \lambda M$, and it is straightforward to check that \eqref{eq:oddNtwogen1}$\sim$\eqref{eq:oddNtwogen4}  are satisfied.  In summary, we have the criteria: 
\paragraph{Stability criteria:} When the 4d spacetime is a spin manifold, a $\Z_4^{\text{em}}$ stable Lagrangian subgroup exists if and only if $N=L^2 M$ and for every choice of bicharacter labeled by $u,v$ with $uv=1$ mod $N$, where $-1$ is a quadratic residue of $M$. We enumerate such $N$'s in Tab.~\ref{tab:4dcriteria}. 
\bigskip

\begin{table}[]
    \centering
    \begin{tabular}{|c|c|c|c|c|c|c|c|c|c|c|c|c|c|c|c|c|c|c|c|
    }
    \hline
 0 & \textcolor{red}{1} & \textcolor{red}{2} & 3 & \textcolor{blue}{4} & \textcolor{red}{5} & 6 & 7 & \textcolor{blue}{8} & \textcolor{blue}{9} & \textcolor{red}{10} & 11 & 12 & \textcolor{red}{13} & 14 & 15 & \textcolor{blue}{16} & \textcolor{red}{17} & \textcolor{blue}{18} & 19
   \\\hline
 \textcolor{blue}{20} & 21 & 22 & 23 & 24 & \textcolor{red}{25} & \textcolor{red}{26} & 27 & 28 & \textcolor{red}{29} & 30 & 31 & \textcolor{blue}{32} & 33 & \textcolor{red}{34} & 35 & \textcolor{blue}{36} & \textcolor{red}{37}
   & 38 & 39 \\\hline
 \textcolor{blue}{40} & \textcolor{red}{41} & 42 & 43 & 44 & \textcolor{blue}{45} & 46 & 47 & 48 & \textcolor{blue}{49} & \textcolor{red}{50} & 51 & \textcolor{blue}{52} & \textcolor{red}{53} & 54 & 55 & 56 & 57
   & \textcolor{red}{58} & 59 \\\hline
 60 & \textcolor{red}{61} & 62 & 63 & \textcolor{blue}{64} & \textcolor{red}{65} & 66 & 67 & \textcolor{blue}{68} & 69 & 70 & 71 & \textcolor{blue}{72} & \textcolor{red}{73} & \textcolor{red}{74} & 75 & 76 & 77
   & 78 & 79 \\\hline
\textcolor{blue}{80} & \textcolor{blue}{81} & \textcolor{red}{82} & 83 & 84 & \textcolor{red}{85} & 86 & 87 & 88 & \textcolor{red}{89} & \textcolor{blue}{90} & 91 & 92 & 93 & 94 & 95 & 96 & \textcolor{red}{97}
   & \textcolor{blue}{98} & 99 \\\hline
\end{tabular}
    \caption{$N$'s that admit $\Z_4^{\text{em}}$ stable Lagrangian subgroups, i.e. $N=L^2 M$ labeled by colors. The red and blue ones are the $N$'s with $L=1$ and $L>1$ respectively. They correspond to the Lagrangian subgroups that are generated by one and two generators. }
    \label{tab:4dcriteria}
\end{table}

Let's make several comments. 
\begin{enumerate}
    \item Combining the above criteria with the general discussion in Sec.~\ref{sec:criterias}, we claim that the duality defect associated with gauging $\Z_N^{(1)}$ is group theoretical if and only if $N=L^2 M$ where $-1$ is a  quadratic residue of $M$.  The if direction is proven in Sec.~\ref{sec:criterias}, and will be further supported by explicitly showing the SymTFT is a Dijkgraaf-Witten theory and also explicitly constructing the topological manipulation for every such $N$. The condition is also independent of the choice of bicharacters $u,v$ and the higher dimensional generalization of the Frobenius-Schur indicator $\epsilon$, which will be further discussed in Sec.~\ref{sec:5dDW}. However the only if direction is less solid, and remains a conjecture. 
    \item On non-spin manifolds, the criteria for odd $N$ remains the same. For even $N$, we don't have the full classification of stable Lagrangian subgroups, but let's comment on two special cases. The first case is when the Lagrangian subgroup is generated by one surface operator, then we are able to show that the stable Lagrangian subgroup does not exist. We present the details in App.~\ref{app:onegen}. The second case is when $N=L^2$ is a perfect square so that $(p,N/p,\ell)=(L,L,0)$ are all even for even $L$. On top of the stability condition on charges \eqref{eq:oddNtwogen1}$\sim$\eqref{eq:oddNtwogen4} whose solution is $(N,x,y,z,w,p,\ell)=(L^2, 0,u,-v,0,L,0)$ for even $L$, there are additional stability conditions on the topological refinements $J_{x,y}=xy+c_1 x +c_2 y$ mod $2$ enforcing $c_1=c_2$ mod $2$. In Sec.~\ref{sec:5dtopomanipulation} we will comment on the explicit topological manipulation on non-spin manifold corresponding to this special case.  
    \item When $L=1$, there is only one generator $\widetilde{S}_{(r, 1)}$, with $(vr)^2= -1$ mod $N$. This is precisely the condition found in \cite[App.~C]{Choi:2022rfe}, as well as in \cite{Bashmakov:2022uek} for prime $N$'s, for the $\Z_N^{(1)}$ duality defect to be group theoretical on spin manifolds. 
    \item In \cite{Apruzzi:2023uma}, the problem of determining group theoretical-ness was phrased in terms of Hanany-Witten transition between strings and 7-branes in the holographic IIB setup. In particular, the main result, summarized in Tab.~5 of  \cite{Apruzzi:2023uma}, coincides with Tab.~\ref{tab:4dcriteria} of the current work for $N\leq 29$. 
    \item It is intriguing to note that the same series $N=L^2 M$ (with $-1$ being a quadratic residue of $M$) was found in a completely different context in \cite{Apte:2022xtu}, where the authors showed that a (spin) TQFT with $\Z_N^{(1)}$  symmetry $N$ satisfies exactly the same condition.  
    In Sec.~\ref{sec:comments}, we explain such a coincidence. 
\end{enumerate}

\subsection{SymTFT as a Dijkgraaf-Witten theory}
\label{sec:5dDW}

In this subsection, we show explicitly that the SymTFT for the entire $\Z_N$ duality symmetry is a Dijkgraaf-Witten theory when the $\Z_4^{\text{em}}$ stable Lagrangian subgroup exists. From Sec.~\ref{sec:3dDW}, we learned that for the SymTFT obtained from gauging the electric-magnetic exchange symmetry of the $\Z_N$ gauge theory to be a Dijkgraaf-Witten theory, we should rewrite the original $\Z_N$ gauge theory in terms of a set of new gauge fields so that the em exchange symmetry acts on the electric and magnetic gauge fields separately, and does not exchange them. We will find below how this can be achieved for 5d $\Z_N$ 2-form gauge theory.

\paragraph{Odd $N$:}
From Sec.~\ref{sec:5dstableLag}, we found that the $\Z_4^{\text{em}}$ stable Lagrangian subgroup is generated by surface operators $\widetilde{S}_{(LM,0)}$ and $\widetilde{S}_{(Lrv, L)}$ with the constraints that $r^2+1=\lambda M$ for some integer $\lambda$. 
Inspired by the discussion in Sec.~\ref{sec:3dDW}, we decompose the $\Z_N$ 2-form gauge fields into new ones as follows
\begin{eqnarray}\label{eq:bbhat}
    (\widehat{b},b) = (LM, 0) \widehat{a} + (Lrv, L) \widehat{c} - (\lambda,0) c - (rv,1) a
\end{eqnarray}
where $(\widehat{a}, \widehat{c}, a,c)$ are $(\Z_L, \Z_{LM},\Z_N,  \Z_{N} )$ cochains respectively. We again label the magnetic fields by hatted letters, while the electric fields by unhatted letters. 
Under $\Z_4^{\text{em}}$, \eqref{eq:Z4em} implies 
\begin{eqnarray}\label{eq:bhatb}
\begin{split}
    (\widehat{b},b) \to (-vb, u\widehat{b}) &= (0, uLM)\widehat{a} + (-vL, Lr)\widehat{c} -(0, u\lambda)c- (-v, r)a\\
    &=[-(LM, 0)r+(Lrv,L)Mu]\widehat{a} + [-(LM,0)\lambda v+(Lrv,L)r]\widehat{c} \\& \quad -[-(\lambda,0)r+ (vr,1)\lambda u]c - [-(\lambda,0) Mv + (v r,1)r]a
\end{split}
\end{eqnarray}
where in the second line we used \eqref{eq:oddNtwogen1}$\sim$\eqref{eq:oddNtwogen4}. Comparing \eqref{eq:bbhat} and \eqref{eq:bhatb}, we find the transformation of new gauge fields as\footnote{The coefficient of $a$ and $c$ in \eqref{eq:bhatb} is designed to ensure that $a,c$ transform in a conjugate way as $\widehat{a}, \widehat{c}$ and also $(\widehat{a},a)$ and $(\widehat{c},c)$ appear as conjugate fields in the Lagrangian. }
\begin{eqnarray}\label{eq:newtrans}
    \begin{split}
        \widehat{a} &\to -r \widehat{a} - \lambda  v \widehat{c},\\
        \widehat{c} &\to u M \widehat{a} + r \widehat{c},\\
        a &\to r a+ u \lambda c,\\
        c&\to -M v a - r c,\\
    \end{split}
\end{eqnarray}
where the electric and magnetic fields do not mix under $\Z_4^{\text{em}}$. Indeed, the generators of the Lagrangian subgroup $e^{\frac{2\pi i p}{N}\oint b} = e^{-\frac{2\pi i}{L}\oint a}$ and $e^{\frac{2\pi i}{N} \oint Lrv b + L\widehat{b}}= e^{-\frac{2\pi i \lambda}{LM}\oint  c }$ are all written in terms of the electric fields, hence their generating set is stable under $\Z_4^{\text{em}}$. 
The Lagrangian of the $\Z_N$ 2-form gauge theory can be rewritten in terms of the new gauge fields as 
\begin{equation}\label{eq:newLag2}
\begin{split}
    \frac{2\pi}{N} \widehat{b} \delta b &= -\frac{2\pi}{L} \widehat{a} \delta a  + \frac{2\pi\lambda}{LM} \widehat{c} \delta c + \frac{2\pi \lambda}{L^2 M} a \delta c + \frac{2\pi}{L^2 M} \delta ( -\lambda L \widehat{c}c + \frac{N+1}{2} L^2 r v\widehat{c}\widehat{c} - Lrv a \widehat{c} + \frac{N+1}{2} rv a a)\\
    &= -\frac{2\pi}{L} \widehat{a} \delta a  + \frac{2\pi\lambda}{LM} \widehat{c} \delta c + \frac{2\pi \lambda}{L^2 M} a \delta c + \CL_{\text{bdy}}.
\end{split}
\end{equation}
In the second line, the first two terms are the BF terms that couple electric fields $(a,c)$ to the magnetic fields $(\widehat{a}, \widehat{c})$, while the third term is the mixed coupling (Dijkgraaf-Witten twist term) that only depends on the electric fields $a,c$.  The last term $\CL_{\text{bdy}}$ is the boundary term which does not play any role in this subsection, while it will be crucial in Sec.~\ref{sec:5dtopomanipulation}. 
Because the $\Z_4^{\text{em}}$ only exchanges among the electric fields and the magnetic fields separately, we conclude that after gauging $\Z_4^{\text{em}}$, the theory is still a Dijkgraaf-Witten. The explicit form the Dijkgraaf-Witten can be analogously constructed as in Sec.~\ref{sec:3dDW}. See also \cite{Kaidi:2022cpf}. Concretely, after gauging $\Z_4^{\text{em}}$, we introduce the $\Z_4^{\text{em}}$ cocycle $x$ as a dynamical gauge field. Denote the two component 2-form gauge field
\begin{eqnarray}
    \mathbf{a}= 
    \begin{pmatrix}
        a\\
        c
    \end{pmatrix}, \qquad 
    \widehat{\mathbf{a}}=
    \begin{pmatrix}
        \widehat{a}\\
        \widehat{c}
    \end{pmatrix}.
\end{eqnarray} 
The gauged $\Z_N$ 2-form gauge theory becomes 
\begin{eqnarray}
\begin{split}
    \CL_{ijklmn}^{\text{gauged}} = &\frac{2\pi}{LM} \widehat{\mathbf{a}}^{T}_{ijk} W K^{x_{ik}} (K^{x_{kl}} \mathbf{a}_{lmn} - \mathbf{a}_{kmn} + \mathbf{a}_{kln} - \mathbf{a}_{klm}) \\
    &+ \frac{\pi \lambda}{N} \mathbf{a}^T_{ijk} V K^{x_{ik}} (K^{x_{kl}} \mathbf{a}_{lmn} - \mathbf{a}_{kmn} + \mathbf{a}_{kln} - \mathbf{a}_{klm}) + \frac{\pi \epsilon}{2} x_{ij} (\beta x)_{jkl}  (\beta x)_{lmn}
\end{split}
\end{eqnarray}
where 
\begin{eqnarray}
    K= 
    \begin{pmatrix}
        r & u\lambda\\
        -Mv & -r
    \end{pmatrix}, \quad
    \widehat{K}= 
    \begin{pmatrix}
        -r & -\lambda v\\
        uM & r
    \end{pmatrix}, \quad
    W= 
    \begin{pmatrix}
        -M & 0\\
        0 & \lambda
    \end{pmatrix}, \quad 
    V=
    \begin{pmatrix}
        0 & 1\\
        -1 & 0
    \end{pmatrix}
\end{eqnarray}
and $(\beta x)_{jkl} := \frac{1}{4}(\delta x)_{jkl} = \frac{1}{4}(x_{kl}-x_{jl}+x_{jk})$ is always an integer valued since $x$ is a $\Z_4$ cocycle. 
The gauged theory is invariant under gauge transformations
\begin{eqnarray}
    \begin{split}
        \mathbf{a}_{ijk} &\to K^{-\gamma_i} (\mathbf{a}_{ijk}+ K^{x_{ij}}\alpha_{jk} - \alpha_{ik} + \alpha_{ij}),\\
        \widehat{\mathbf{a}}_{ijk} & \to \widehat{K}^{-\gamma_i} (\widehat{\mathbf{a}}_{ijk}+ \widehat{K}^{x_{ij}}\widehat{\alpha}_{jk} - \widehat{\alpha}_{ik} + \widehat{\alpha}_{ij}),\\
        x_{ij} &\to x_{ij} + \gamma_j - \gamma_i,
    \end{split}
\end{eqnarray}
thanks to the identities $\widehat{K}^{T} W K = W$ and $K^T V K = V$. 
Summing over $\widehat{\mathbf{a}}$ constrains $\mathbf{a}$ to be a twisted cocycle, i.e. $\mathbf{a}$ is twisted-flat. In components, $W(K^{x_{kl}} \mathbf{a}_{lmn} - \mathbf{a}_{kmn} + \mathbf{a}_{kln} - \mathbf{a}_{klm})=0$ mod $LM$.  This means that $\mathbf{a},x$ form a non-trivial 2-group. The full partition function is 
\begin{equation}
    \CZ_{\text{gauged}} = \sum_{\mathbf{a},x} \prod_{ijklmn} \exp\left( \frac{ i\pi \lambda}{N} \mathbf{a}^T_{ijk} V K^{x_{ik}} (K^{x_{kl}} \mathbf{a}_{lmn} - \mathbf{a}_{kmn} + \mathbf{a}_{kln} - \mathbf{a}_{klm}) + \frac{i \pi \epsilon}{2} x_{ij} (\beta x)_{jkl}  (\beta x)_{lmn} \right)
\end{equation}
where we only sum over the gauge fields $\mathbf{a},x$ with the twisted-flatness condition. 
The partition function is obviously of the form of Dijkgraaf-Witten (see Eq.~\eqref{eq:DWdef}).

\paragraph{Even $N$:}
When the 4d spacetime is a spin manifold, the discussion is almost the same as above, such as \eqref{eq:bbhat}, \eqref{eq:bhatb} and \eqref{eq:newtrans}. The only modification is to replace the factor $\frac{N+1}{2}$ in the Lagrangian \eqref{eq:newLag2} by $\frac{1}{2}$,  and the self pairing e.g. $aa$ by the Pontryagin square of $a$, i.e. $\CP(a)$. When the 4d spacetime is a non-spin manifold, in the special case where the Lagrangian subgroup has only a single generator, we find in App.~\ref{app:onegen} that there is no stable Lagrangian subgroup, hence the SymTFT obtained by gauging $\Z_4^{\text{em}}$ can not be a Dijkgraaf-Witten. When there are two generators, we do not solve the stability condition in this paper.

\bigskip

In summary, for all the cases where the stable Lagrangian subgroup exist, we showed that the SymTFT for the duality symmetry is a Dijkgraaf-Witten theory.

\subsection{Explicit topological manipulations}
\label{sec:5dtopomanipulation}

We finally proceed to find explicit topological manipulations that map the duality defect to an invertible defect.

Suppose a 4d QFT with $\Z_N^{(1)}$ one-form symmetry is self-dual under gauging $\Z_N^{(1)}$. This means 
\begin{eqnarray}\label{eq:selfduality5d}
    \CZ_{\CX}[B] = \sum_{b\in \Z_N} \CZ_{\CX}[b] e^{-\frac{2\pi i}{N}\int_{M_4} v bB}.
\end{eqnarray}
The summation $b\in \Z_N$ means summing over 2-cochain $b$ valued in $\Z_N$ with flatness condition $\delta b= 0$ mod $N$, modulo gauge transformations. The idea of identifying the topological manipulation which maps the duality defect to an invertible defect is the same as in Sec.~\ref{sec:3dtopomanipulation}.  The self-dual theory $\CX$ corresponds to Dirichlet boundary condition of $b$. The desired topological manipulation should map such boundary condition to a new topological boundary condition where all the new electric fields have Dirichlet boundary conditions. In the meantime, the discrete theta terms should be introduced coming from the boundary terms of the SymTFT $\CL_{\text{bdy}}$ in \eqref{eq:newLag2}.

\paragraph{Odd $N$:}
In this case, $N$ has the form $N=L^2 M$ where $-1$ is a quadratic residue of $M$. The topological manipulation should map the Dirichlet boundary condition of $b$ to the Dirichlet boundary conditions of $a,c$ defined via \eqref{eq:bbhat}. Denoting the background field for the former as $B$, and those for the latter as $A$ and $C$, we have  
\begin{eqnarray}\label{eq:5dodd3}
    \bra{B} \to \sum_{\widehat{c}\in \Z_{LM}} \bra{L\widehat{c}-A} e^{-\frac{2\pi i}{LM}\int_{M_4} (\lambda C+ rv A) \widehat{c} + \frac{2\pi i}{N} \frac{N+1}{2} \int_{M_4} (L^2 rv \widehat{c}\widehat{c}+ rv AA)}.
\end{eqnarray}
The phase factor on the RHS is the $\CL_{\text{bdy}}$ in \eqref{eq:newLag2}. 
This implies that the new partition function obtained by topological manipulation is
\begin{equation}\label{eq:445}
    \CZ_{\widetilde{\CX}}[A,C] = \sum_{\widehat{c}\in \Z_{LM}} \CZ_{\CX}[L\widehat{c}-A]e^{-\frac{2\pi i}{LM}\int_{M_4} (\lambda C+ rv A) \widehat{c} + \frac{2\pi i}{N} \frac{N+1}{2} \int_{M_4} (L^2 rv \widehat{c}\widehat{c}+ rv AA)}.
\end{equation}
Notice that we start with taking the background gauge field $A,C$ to be $\doubleZ_N$ gauge field for convenience. However, the action \eqref{eq:445} is invariant up to terms containing only background fields under the following transformation:
\begin{equation}\label{eq:twogeneratorgauge}
\begin{split}
    A \rightarrow A + LX,  \qquad
     C \rightarrow C+LMY, \qquad
     \widehat{c} \rightarrow \widehat{c} +X,    
\end{split}
\end{equation}
where $X,Y \in \doubleZ_N$.
Eq.~\eqref{eq:twogeneratorgauge} shows that $A$ is actually a $\doubleZ_{L}$ background field and $C$ is a $\doubleZ_{LM}$ background field. 

We proceed to check what does the duality symmetry of $\CX$ is mapped to in $\widetilde{\CX}$. Substituting \eqref{eq:selfduality5d} into \eqref{eq:445}, we find 
\begin{equation}\label{eq:447}
    \begin{split}
        \CZ_{\widetilde{\CX}}[A,C] = \sum_{\widehat{c}\in \Z_{LM}, b\in \Z_N} \CZ_{\CX}[b]e^{-\frac{2\pi i}{N}\int_{M_4} vb(L\widehat{c}-A)-\frac{2\pi i}{LM}\int_{M_4} (\lambda C+ rv A) \widehat{c} + \frac{2\pi i}{N} \frac{N+1}{2} \int_{M_4} (L^2 rv \widehat{c}\widehat{c}+ rv AA)}.
    \end{split}
\end{equation}
We first sum over $\widehat{c}$, where the relevant terms are $\sum_{\widehat{c}\in \Z_{LM}} e^{-\frac{2\pi i}{LM} \int_{M_4} (vb+\lambda C + rvA)\widehat{c} + \frac{2\pi i}{M}\frac{N+1}{2}rv \int_{M_4} \widehat{c}\widehat{c}}$. This sum is non-vanishing only when $vb+\lambda C + rvA$ can be divided by $L$. \footnote{This can be seen by replacing $\widehat{c}$ with $\widehat{c}+s$ where $s$ is an integer valued cochain, and demand the sum does not change as $\widehat{c}$ is a dummy variable. } We thus define $b=Lb' - u(\lambda C + rv A)$. Substituting this into \eqref{eq:447} and summing over $\widehat{c}$, we get
\begin{equation}\label{eq:5dodd4}
    \begin{split}
        \CZ_{\widetilde{\CX}}[A,C] &= \sum_{\widehat{c}, b'\in \Z_{LM}} \CZ_{\CX}[Lb'-u(\lambda C + rv A)] e^{-\frac{2\pi i}{N}\int_{M_4}(-vLb'+ \lambda C + rv A)A + \frac{2\pi i}{N}\frac{N+1}{2}\int_{M_4} rv AA + L^2 rv b'b'}\\
        &= \CZ_{\widetilde{\CX}}[u\lambda C+ rA, -v M A - r C] e^{-\frac{2\pi i}{N}\frac{N+1}{2} \lambda\int_{M_4} (2\lambda MAC+ ru\lambda CC+ rv M AA)}.
    \end{split}
\end{equation}
Therefore we have shown that the duality symmetry of $\CX$ becomes an invertible symmetry mapping on the background fields of the new theory $\widetilde{\CX}$ as 
\begin{eqnarray}\label{eq:5dodd5}
    A\to u \lambda C+rA \text{ mod } L, \qquad C\to -vMA-rC \text{ mod } ML.
\end{eqnarray}
This agrees with \eqref{eq:newtrans}.

\paragraph{Even $N$:}
When the 4d spacetime is a spin manifold, the discussion is almost identical to the odd $N$ case. The only modifications are to replace $\frac{N+1}{2}$ by $\frac{1}{2}$, and self pairing, e.g. $BB$, by $\CP(B)$ in \eqref{eq:5dodd3}, \eqref{eq:445}, \eqref{eq:447}, \eqref{eq:5dodd4}. The duality symmetry \eqref{eq:selfduality5d} is mapped to an invertible symmetry \eqref{eq:5dodd5}.

When the 4d spacetime is a non-spin manifold, we haven't classified all the solutions. However, as mentioned in Sec.~\ref{sec:5dstableLag}, when $N=L^2$ for even $L$, there is a stable Lagrangian subgroup. Inspired by a similar discussion in Sec.~\ref{sec:3dtopomanipulation}, we claim that the topological manipulation is simply gauging the $\Z_L^{(1)}$ subgroup, i.e. $\widetilde{\CX}=\CX/\Z_L^{(1)}$. Concretely, 
\begin{eqnarray}\label{eq:nonspinZ}
    \CZ_{\widetilde{\CX}}[A,C]= \sum_{a\in \Z_L} \CZ_{\CX}[L \widehat{c}- A] e^{-\frac{2\pi i}{L}\int_{X_4} \widehat{c}C}
\end{eqnarray}
where $A,C$ are both $\Z_L$ 2-form gauge fields. Substituting \eqref{eq:selfduality5d} into \eqref{eq:nonspinZ}, we find the background fields $A,C$ are mapped as
\begin{eqnarray}
    A\to uC \text{ mod }L, \qquad C\to -vA \text{ mod }L~.
\end{eqnarray}
Hence when the 4d spacetime is a non-spin manifold, we have at least confirmed a spacial case where $N=L^2$ such that the $\Z_N^{(1)}$ duality defect is group theoretical. 

\bigskip

In summary, for all the cases where the stable Lagrangian subgroup exist (on spin manifolds), we have constructed explicit topological manipulation which maps the duality symmetry to an invertible symmetry.

\section{Connection with obstruction to duality-preserving gapped phases}
\label{sec:comments}

We have observed in the above that for the $\Z_N^{(d/2-1)}$ duality defect to be group theoretical, some number-theoretic condition should be satisfied, and exactly the same condition has appeared in other contexts, including whether there exists an SPT or TQFT invariant under gauging $\Z_N^{(d/2-1)}$\cite{Choi:2021kmx, Choi:2022zal, Apte:2022xtu, Zhang:2023wlu}. In this section, we comment on the relation between them. See also \cite{Antinucci:2023ezl,Cordova:2023bja} for related discussions.

\subsection{Duality defects, gauging, and SPT}
\label{sec:SPT}

Given a finite abelian group $\Z_N^{(d/2-1)}$ and an integer $u$ coprime with $N$ which specifies how to gauge $\Z_N^{(d/2-1)}$, one may look for $\Z_N^{(d/2-1)}$ SPTs in $d$ dimensions satisfying the one of the following closely related properties:
\begin{enumerate}
    \item[(a)]  The $\Z_N^{(d/2-1)}$ SPT is equipped with a $\Z_N^{(d/2-1)}$ duality defect with an \emph{unspecified} Frobenius-Schur indicator (labeled by $\epsilon$). Note that the bicharacter is specified by $u$. Equivalently, the $\Z_N^{(d/2-1)}$ SPT is invariant under gauging $\Z_N^{(d/2-1)}$;
    \item[(b)]  The $\Z_N^{(d/2-1)}$ SPT is equipped with a $\Z_N^{(d/2-1)}$ duality defect with a \emph{given} Frobenius-Schur indicator (labeled by $\epsilon$). 
\end{enumerate}
A $\Z_N^{(d/2-1)}$-SPT satisfying property (b) automatically satisfies property (a) since $\Z_N^{(d/2-1)}$ duality defect implements gauging $\Z_N^{(d/2-1)}$. However, the converse is not true: it is possible that for a given coprime pair $(N,u)$, the duality defect with certain $\epsilon$ may not be realized by any $\Z_N^{(d/2-1)}$ SPT, or equivalently, the duality defect may be anomalous for some choice of $\epsilon$. See \cite{Thorngren:2019iar} for the general criteria of the anomaly in 2d.

In \cite{Choi:2021kmx, Choi:2022zal, Apte:2022xtu, Kaidi:2022cpf, Thorngren:2019iar}, the authors discussed the $G$-SPTs with property (a). It was found that 
\begin{enumerate}
    \item in 2d and among all $N$'s, the only $\Z_N^{(0)}$ SPT is a trivial SPT with partition function $\CZ=1$. Furthermore, the trivial SPT is not invariant under gauging $\Z_N^{(0)}$ for any coprime pair $(N, u)$;
    \item in 4d and among all $N$'s, an $\Z_N^{(1)}$ SPT is classified by
    \begin{eqnarray}
    \begin{split}
        \frac{2\pi k}{N}\frac{N+1}{2} \int_{X_4} B^{(2)}B^{(2)}, &\quad k\in \Z_N, \quad X_4=\text{ spin or non-spin},\\
        \frac{2\pi k}{2N}\int_{X_4} \CP(B^{(2)}), &\quad
        \begin{cases}
            k\in \Z_{2N}, &\quad X_4=\text{ 
 non-spin},\\
            k\in \Z_N, &\quad X_4=\text{ spin},\\
        \end{cases}
    \end{split}
    \end{eqnarray}
    and there exists an $\Z_N^{(1)}$ SPT invariant under gauging $\Z_N^{(1)}$ for a given $u$ if and only if there is an integer $r$ solving the following equation\footnote{In \cite{Choi:2021kmx} only the case $u=1$ was discussed. But it is straightforward to check that the same condition holds for any $u$ coprime with $N$. }
    \begin{eqnarray}
    r^2=-1 \text{ mod } 
    \begin{cases}
        N, & \text{ odd } N \text{ on spin and non-spin $X_4$}, \\
        N, & \text{ even } N \text{ on spin } X_4,\\
        2N, & \text{ even } N \text{ on non-spin } X_4.
    \end{cases}
\end{eqnarray}
\end{enumerate}

Solving (b) is harder. In 2d, for general $G$, a classification of $G$ SPTs equipped with a $G$ duality defect was achieved in \cite{Thorngren:2019iar}, see also \cite{Zhang:2023wlu}. In 3d, some related recent works on the anomaly of fusion 2-categories can be found in \cite{Decoppet:2022dnz, Decoppet:2023bay}.\footnote{The condition for (b) was later determined in \cite{Antinucci:2023ezl,Cordova:2023bja}.}

\subsection{Duality defects, gauging, TQFT, and relation with group theoretical duality defects}
\label{sec:TQFT}

Given an abelian group $\Z_N^{(1)}$ and an integer $u$ coprime with $N$ which specifies how to gauge $\Z_N^{(1)}$, one may look for $\Z_N^{(1)}$ symmetric 4d TQFTs\footnote{In this subsection, we constrain our discussion to duality defects in 4d. The reason is that in 2d, the TQFTs with one ground state on $S^1$ spatial manifold is always an SPT, hence the discussion reduces to Sec.~\ref{sec:SPT}. } satisfying the one of the following closely related properties:
\begin{enumerate}
    \item[(a')]  The $\Z_N^{(1)}$ symmetric TQFT (with one ground state on $S^3$ spatial manifold) is equipped with a $\Z_N^{(1)}$ duality defect with an \emph{unspecified}  
    Frobenius-Schur indicator (labeled by $\epsilon$). Note that the bicharacter is specified by $u$. Equivalently, the $\Z_N^{(1)}$ symmetric TQFT is invariant under gauging $\Z_N^{(1)}$; 
    \item[(b')]  The $\Z_N^{(1)}$ symmetric TQFT (with one ground state on $S^3$ spatial manifold) is equipped with a $\Z_N^{(1)}$ duality defect with a \emph{given} 
    Frobenius-Schur indicator (labeled by $\epsilon$). 
\end{enumerate}
The requirement of one ground state on $S^3$ spatial manifold ensures that the $\Z_N^{(1)}$ duality symmetry is not spontaneously broken, however, the $\Z_N^{(1)}$ symmetry is allowed to be spontaneously broken since we don't require one ground state on other spatial manifolds such as $T^3$.
Similar to the discussion in Sec.~\ref{sec:SPT}, a $\Z_N^{(1)}$ symmetric TQFT satisfying property (b') automatically satisfies property (a'). However, the converse is not true.

In \cite{Apte:2022xtu}, a classification for (a'), when the 4d manifold is spin, has been achieved. It was found that a 4d (spin) TQFT is invariant under gauging $\Z_N^{(1)}$ if and only if $N$ has the form 
\begin{eqnarray}\label{eq:obsfree}
    N=L^2 M, \qquad \text{$L\in \Z$, and $\exists$ an integer $r$ solving } r^2= -1 \text{ mod } M.
\end{eqnarray}
This is precisely the condition for the $\Z_N^{(1)}$ duality defects to be group theoretical.

\begin{figure}
    \centering
    \begin{tikzpicture}
        \filldraw[color=red!60, fill=red!0, very thick] (0,0) ellipse (2 and 1);
        \filldraw[color=orange!60, fill=orange!5, very thick, fill opacity=0] (0,-0.8) ellipse (3.5 and 3);
        \node[] at (0,0) {\begin{tabular}{c}
                $\exists ~\Z_N^{(0)}$SPT \\invariant
                under gauging\\
                (a)
        \end{tabular}};
    \node[] at (0,-2) {\begin{tabular}{c}
            Group theoretical\\
            $\Z_N^{(0)}$ duality defect
    \end{tabular}};
    \end{tikzpicture}
    \caption{For 2d theory: space of $N$ where group theoretical $\Z_N^{(0)}$ duality defects exist (bounded by orange circle), and where $\Z_N^{(0)}$ duality defect is anomaly free (bounded by blue circle). The orange circle bounds the regime where $N$ is a perfect square, and the blue circle bounds only a point $N=1$. As a consequence, the cases satisfying (b) also only contains a trivial point $N=1$. This further implies that the condition for (b) to be satisfied is also $N=1$ since (b) is more restricted than (a).}
    \label{fig:Venn2d}
\end{figure}

\begin{figure}
    \centering
    \begin{tikzpicture}
        \filldraw[color=blue!60, fill=blue!5, very thick] (0,0) ellipse (2 and 1);
        \filldraw[color=red!60, fill=red!5, very thick, fill opacity=0] (0,-0.5) ellipse (4.2 and 2);
        \filldraw[color=orange!60, fill=orange!5, very thick, fill opacity=0] (0,-1.3) ellipse (6 and 3.5);
        \node[] at (0,0) {\begin{tabular}{c}
                Anomaly free \\duality defect\\
                (b)
        \end{tabular}};
        \node[] at (0,-1.7) {\begin{tabular}{c}
                $\exists ~\Z_N^{(1)}$SPT invariant
                under gauging\\
                (a)
        \end{tabular}};
    \node[] at (0,-3.3) {
    \begin{tabular}{c}
         $\exists ~\Z_N^{(1)}$TQFT invariant\\
                under gauging\\
                (a')
    \end{tabular}
    $=$
    \begin{tabular}{c}
            Group theoretical\\
            $\Z_N^{(1)}$ duality defect
    \end{tabular}};
    \end{tikzpicture}
    \caption{For 4d theory: space of $N$ where group theoretical $\Z_N^{(1)}$ duality defects exist (bounded by orange circle) which coincides with the situation where there exists a $\Z_N^{(1)}$-TQFT invariant under gauging $\Z_N^{(1)}$ (hence also bounded by orange circle), where there exists a $\Z_N^{(1)}$ SPT invariant under gauging $\Z_N^{(1)}$ (bounded by red circle), and where the $\Z_N^{(1)}$ duality defect is anomaly free (bounded by blue circle). The orange circle bounds the regime where $N$ takes the form $L^2 M$ and $-1$ is a quadratic residue of $M$. The red circle bounds the regime where $-1$ is a quadratic residue of $N$. The condition for the blue circle depends on the choice of the (higher dimensional generalization of) FS indicator, which is not determined by this current work, but is later determined in \cite{Antinucci:2023ezl,Cordova:2023bja}. }
    \label{fig:Venn4d}
\end{figure}

To understand why there is such a coincidence, we again use the SymTFT. 
It turns out that the SymTFT is a  natural set up in discussing 4d TQFTs with the property (a') or (b'). Any 4d TQFT with a non-anomalous $\Z_N^{(1)}$ global symmetry can be expanded into a 5d slab where in the 5d bulk is a $\Z_N$ 2-form gauge theory \eqref{eq:leftbdy}, the left boundary is the Dirichlet boundary condition \eqref{eq:rightbdy}, and the right boundary is another topological boundary condition  \eqref{eq:dynbdystate} specified by the 4d TQFT. Since the SymTFT does not contain genuinue line or point operator, it is clear that after shrinking the 5d slab to get a genuine 4d theory, there isn't any non-trivial topological local operator, neither directly coming from point operator in the bulk nor from compactifying line operators along the shrinked direction. Hence the only topological local operator is the trivial identity, and the unique ground state on $S^3$ is guaranteed. Furthermore, it is also known \cite{Kaidi:2022cpf, Freed:2022qnc,Bhardwaj:2023ayw, Zhang:2023wlu, Antinucci:2022vyk} that gauging $\Z_N^{(1)}$ of the 4d TQFT can be equivalently achieved by fusing a $\Z_4^{\text{em}}$ symmetry defect (constructed as a condensation defect in \cite{Kaidi:2022cpf, Antinucci:2022vyk}) of the 5d $\Z_N^{(1)}$ gauge theory to either of the topological boundary. Requiring (a'), i.e. the TQFT to be invariant under gauging $\Z_N^{(1)}$, amounts to requiring the right topological boundary state \eqref{eq:dynbdystate} to be invariant under fusing with the $\Z_4^{\text{em}}$ symmetry defect. This is precisely the stability condition in Sec.~\ref{sec:5dstableLag}, from where we derived the same condition as \eqref{eq:obsfree}.

Solving (b') is again harder. To achieve a full classification of TQFTs equipped with a duality defect, we need to consider the SymTFT of the duality symmetry, i.e. $\Z_N^{(1)}$ 2-form gauge theory with $\Z_4^{\text{em}}$ gauged, as in \eqref{eq:symtft}. Note that the resulting SymTFT depends on both the choice of bicharacters (see the discussion below \eqref{eq:symtft}), as well as the choice of a discrete theta term of the $\Z_4^{\text{em}}$ gauge field $x$. The choice of discrete theta term can be understood as a higher dimensional generalization of the Frobenius Schur indicator as discussed in Sec.~\ref{sec:5dDW}.  There is a topological line operator $K= e^{\frac{i \pi}{2} \oint x}$, which by construction topologically terminates on the left topological boundary, just as the electric surface operator $e^{\frac{2\pi i}{N}\oint b}$ does. To ensure that the duality symmetry is not spontaneously broken (or equivalently there is one ground state on $S^3$), we should demand that the $K$-line can not topologically terminate on the right topological boundary. This imposes additional constraints for (b'), which will be left for future study.

We summarize the main results of $\Z_N^{(0)}$ duality defects in 2d in Fig.~\ref{fig:Venn2d}, and those of $\Z_N^{(1)}$ duality defects in 4d in Fig.~\ref{fig:Venn4d}. \footnote{We thank Philip Boyle Smith for the discussions on the Venn diagrams. }

\section*{Acknowledgements}
We are grateful to Justin Kaidi for collaboration at the early stage of the project. We also thank Philip Boyle Smith, Justin Kaidi, and Kantaro Ohmori for useful discussions, and John McGreevy, Ken Intriligator, Justin Kaidi, Kantaro Ohmori, and Carolyn Zhang for useful comments on the draft.  Z.S. is supported from the US Department of Energy (DOE) under cooperative research agreement DE-SC0009919, Simons Foundation award No. 568420 (K.I.) and the Simons Collaboration on Global Categorical Symmetries. Y.Z. is partially supported by WPI Initiative, MEXT, Japan at IPMU, the University of Tokyo.

\appendix

\section{Lagrangian subgroup for $\doubleZ_N$ $2$-form  gauge theory in 5d}
\label{app:alternative}

In this appendix, we provide an alternative derivation of the Lagrangian subgroups for $\doubleZ_N$ $2$-form 
gauge theory in $5$d. 
Recall the Lagrangian subgroup consists of  maximal number of pairs $(e,m)$ such that any two pairs $(e,m)$ and $(e',m')$ satisfy $e^{-\frac{2\pi i}{N}(em'-me')\langle \sigma,\sigma'\rangle}=1$. 
The maximal condition ensures that if we condense all the surface operators in the Lagrangian subgroup, every surface operator outside the Lagrangian subgroup braids non-trivially with at least one surface operator in the Lagrangian subgroup, and hence is projected out.

To proceed, we quote a theorem from \cite[Sec.~4.2]{tignol1986symplectic}, which claims that the Lagrangian subgroups are classified by a subgroup $Q$ of $H = \doubleZ_N$ and a symmetric bilinear form $\Psi$ on $Q$, that is, $\Psi: Q\times Q \rightarrow U(1)$ where
\begin{equation}\label{eq:sym_bilinear_form}
    \Psi(h_1,h_2) = \Psi(h_2, h_1), \quad \Psi(h_1 h_2, h_3) = \Psi(h_1, h_3) \Psi(h_2, h_3).
\end{equation}
In this case, $Q = \doubleZ_p$ for some integer $p$ dividing $N$.  To see all symmetric bilinear forms $\Psi$ on $\doubleZ_p$, let $\eta$ denote the generator of $\doubleZ_p$, and from \eqref{eq:sym_bilinear_form} we find
\begin{equation}
    \Psi(\eta^m,\eta^n) = \Psi(\eta, \eta)^{mn}.
\end{equation}
Hence, $\Psi$ is completely determined by $\Psi(\eta,\eta)$.  First note that $\Psi(\eta, 1)= \Psi(1, \eta)=1$.\footnote{This follows from the second condition in \eqref{eq:sym_bilinear_form}, where $h_1=h_2=1$ and $h_3=\eta$. } Taking $m = p$ and $n = 1$ in the above equation, we find $\Psi(\eta,\eta) = e^{\frac{2\pi i \ell}{p}}$ where $\ell = 0,1,\cdots,p-1$. Hence, we will use the $\Psi_{p,\ell}$ to denote the symmetric bilinear form such that 
\begin{equation}\label{eq:Psipl}
    \Psi_{p,\ell}(\eta,\eta) = e^{\frac{2\pi i}{p} \ell}.   
\end{equation}

The Ref.~\cite[Sec.~4.2]{tignol1986symplectic} also specifies how the elements in the Lagrangian subgroup can be constructed from $(Q, \Psi_{p,\ell})$. Denote an arbitrary operator with electric and magnetic charge $(x,y)$ as $\alpha^x \beta^y$, with $y\in \Z_N$ and $x\in \Hom(\Z_N, U(1))\simeq \Z_N$. We also define the standard pairing $\alpha(\beta)=e^{\frac{2\pi i}{N}}$. The key statement is that $\alpha^x \beta^{y}$ belongs to the Lagrangian subgroup specified by $(Q, \Psi_{p,\ell})$ if 
\begin{itemize}
    \item the magnetic charge takes value in $Q$, i.e. $y= \frac{N}{p} y'$ with $y'\in Q$;
    \item the electric charge is constrained by the pairing relation: $\alpha^z(\beta^y) = \Psi_{p, \ell}(b^{y}, b^{z})$ for any $z= \frac{N}{p}z'$ with $z'\in Q=\Z_p$. 
\end{itemize}
By definition, $\alpha^z(\beta^y)= e^{\frac{2\pi i}{N}\frac{N}{p} z' x}$. Using \eqref{eq:Psipl}, the above condition gives
\begin{eqnarray}
    z'(x-\ell y') = 0 \text{ mod }p, \qquad \forall z'\in \Z_p.
\end{eqnarray}
This enforces $x= \ell y' + p x'$ with $x'\in \Z_{N/p}$. Thus the charges are
\begin{eqnarray}
    (x,y) = (\ell y'+ px', y' N/p ) = x'(p,0) + y'(\ell, N/p), \qquad x'\in \Z_{N/p}, \quad y'\in \Z_p.
\end{eqnarray}
In other words, the charges in the Lagrangian subgroup are generated by 
\begin{equation}
    \left(\ell, N/p\right), \quad \left(p,0\right).
\end{equation}
Notice that the two generators could be linearly dependent in general, but they nevertheless generated the full Lagrangian algebra.

\section{Lagrangian subgroup with one generator}
\label{app:onegen}

In this appendix, we focus on the case where the Lagrangian subgroup of 5d $\Z_N$ 2-form gauge theory is generated by a single surface operator $\widetilde{S}_{(p,q)}$, with the coprime condition $\gcd(p,q,N)=1$. This special case has been explored in \cite{Chen:2021xuc}. 
\begin{eqnarray}\label{eq:onegenA}
    \CA_{(p,q)}=\{\widetilde{S}_{(kp,kq)}|k\in \Z_N,\gcd(p,q,N)=1\}.
\end{eqnarray}
Because of the coprime condition, it generates $N$ distinct operators $\widetilde{S}_{(kp,kq)}$ with $k=0,1,..., N-1$.  The trivial mutual braiding condition is clearly satisfied. Different pairs $(p,q)$ may generate the same Lagrangian subgroup, for instance when $N=5$, $(p,q)=(1,1)$ and $(p',q')=(3,3)$ generate the same Lagrangian subgroup $\{\widetilde{S}_{(k,k)}| k\in \Z_5\}$. Such redundancy will not be a problem for our purposes. 

To determine the counter term within $\widetilde{S}_{(kp,kq)}$, we
substitute $(e,m)=k(p,q)$ in \eqref{eq:fusion1}, and find
\begin{eqnarray}\label{eq:Kpkq}
    K_{kp,kq}= 
    \begin{cases}
        \frac{2\pi}{N} \frac{N+1}{2} k^2 pq \text{ mod } 2\pi, & \text{ odd } N \\
        \frac{2\pi}{2N} k^2 pq + \pi J_k \text{ mod } 2\pi, &  \text{ even } N
    \end{cases}.
\end{eqnarray}
Substituting \eqref{eq:Kpkq} into \eqref{eq:fusion2}, we find that for odd $N$ \eqref{eq:fusion2} is automatically satisfied, while for even $N$, $J_k$ should satisfy
\begin{eqnarray}
    J_k= kJ \text{ mod }2
\end{eqnarray}
where $J=0$ or $1$. These two solutions are precisely the topological refinement discussed extensively in \cite{Chen:2021xuc}.

For even $N$, it is useful to extend the range of charges from $\Z_N$ to $\Z_{2N}$, so that shifting the topological refinement $J\to J+1$ can be replaced by shifting the electric or magnetic charge by $N$. For simplicity, take $k=1$. Since $\gcd(p,q,N)=1$, $p,q$ can not be both even. Suppose $p$ is odd. Then $J\to J+1$ can be achieved by shifting $q\to q+N$.

In summary, when there is a single generator, the Lagrangian subgroup of 5d $\Z_N$ 2-form gauge theory is generated by the surface operator
\begin{eqnarray}
    \widetilde{S}_{(p,q)}(\sigma)= 
    \begin{cases}
        S_{(p,0)}(\sigma) S_{(0,q)}(\sigma) e^{\frac{2\pi i}{N}\frac{N+1}{2} pq \braket{\sigma,\sigma}}, & \text{ odd } N\\
        S_{(p,0)}(\sigma) S_{(0,q)}(\sigma) e^{\frac{2\pi i}{2N} pq \CP(\sigma) + i \pi J \braket{\sigma,\sigma}}, & \text{ even } N\\
    \end{cases}
\end{eqnarray}
where $J=0,1$ specifies the topological refinement in \cite{Chen:2021xuc} on 4d non-spin spacetime manifold. 
When the 4d spacetime manifold is spin, the $J$ dependence is trivialized.

\paragraph{$\Z_4^{\text{em}}$ stable Lagrangian subgroup for odd $N$:}
Under $\Z_4^{\text{em}}$, it is mapped to $\widetilde{S}_{(-vq,up)}(\sigma)$, which should also belong to the Lagrangian subalgebra, if stable. So stable Lagrangian subalgebra implies that there exists $k$, such that
\begin{eqnarray}\label{eq:oddNonegen}
    k (p,q)= (-vq,up) \mod N.
\end{eqnarray}
Given \eqref{eq:oddNonegen}, the $\Z_4^{\text{em}}$ image of any other element $a(p,q)$ is also within the Lagrangian subalgebra, $a(-vq, up) = ka(p, q) \mod N$. So the Lagrangian subalgebra is $\Z_4^{\text{em}}$ stable if and only if \eqref{eq:oddNonegen} is satisfied.

For which $N, u, v$ do there exist $k,p,q$ such that \eqref{eq:oddNonegen} holds? We first note that \eqref{eq:oddNonegen} implies  
\begin{eqnarray}\label{eq:F17}
    (k^2+1)p =0 \mod N, \qquad (k^2+1)q= 0 \mod N.
\end{eqnarray}
Further combining with $\gcd(p,q,N)=1$, we have $xp+yq=1\mod N$. Multiplying $x$ and $y$ to the above two equations in \eqref{eq:F17}, we find $k^2=-1 \mod N$. Conversely, given $k$ satisfying $k^2=-1 \mod N$, we simply take $(p,q)=(vk,u)$ such that  $\gcd(p,q,N)=1$ and \eqref{eq:oddNonegen} is satisfied. Thus we have shown that \eqref{eq:oddNonegen} holds if and only if 
\begin{eqnarray}\label{eq:oneNcond}
    k^2+1=0 \mod N
\end{eqnarray}
holds, for any $u,v$. This corresponds to the special case  $L=1$ in Sec.~\ref{sec:5dstableLag}.
As commented there, \eqref{eq:oneNcond} is precisely the condition where the $\Z_N^{(1)}$ duality defect in 4d with odd $N$ can be mapped to an invertible defect discussed in \cite[App.C]{Choi:2022zal}, and also in \cite{Bashmakov:2022uek} for prime $N$. Such odd $N$'s belong to the red series listed in Tab.~\ref{tab:4dcriteria}.

\paragraph{$\Z_4^{\text{em}}$ stable Lagrangian subgroup for even $N$:}
As pointed out in Sec.~\ref{sec:Lag2}, the generator also depends on the choice of topological refinement $J=0,1$, and different choices can be packaged by extending the range of electric and magnetic charges from $\Z_{N}$ to $\Z_{2N}$, hence we take $p,q\in \Z_{2N}$ below for convenience. Under $\Z_4^{\text{em}}$, the generator $\widetilde{S}_{(p,q)}(\sigma)$ is mapped to 
\begin{eqnarray}
    \widetilde{S}_{(p,q)}(\sigma) \to \widetilde{S}_{(-vq,up)}(\sigma),
\end{eqnarray}
and the Lagrangian subgroup is $\Z_4^{\text{em}}$ stable if and only if there exists $k$ such that
\begin{eqnarray}\label{eq:F25}
    (-vq,up) = k(p,q) \mod 2N.
\end{eqnarray}
By applying the same argument as for \eqref{eq:oneNcond}, we find that \eqref{eq:F25} is again equivalent to 
\begin{eqnarray}
    k^2=-1 \mod 2N.
\end{eqnarray}
There is no solution to this for any even $N$. However, if we restrict to 4d spin manifolds,  different topological refinements are trivialized and the electric and magnetic charges obey $p\sim p+N$ and $q\sim q+N$, i.e. the charges are back to $\Z_N$ valued.  Thus \eqref{eq:F25} reduces to $(-vq,up) = k(p,q) \mod N$, which is equivalent to $k^2=-1 \mod N$. This again reproduces the special case $L=1$ in  Sec.~\ref{sec:5dstableLag} as well as the results in \cite[App.~C]{Choi:2022zal}.

\bibliographystyle{ytphys} 
\baselineskip=.95\baselineskip
\bibliography{ref}

\end{document}